\documentstyle[12pt,engine,A4,PSFigure,epsf]{article}
\newcommand{\vp}{\vec{p}}

\font\textf  = cmr12          
\font\absf   = cmr11          
\font\titf   = cmr16          
\font\btitf  = cmr24          

\title{\btitf Artificial Neural Networks for Solving Ordinary and
Partial Differential Equations} 

\author{\titf I. E. Lagaris, A. Likas and D. I. Fotiadis \\
Department of Computer Science \\
University of Ioannina \\
P.O. Box 1186 - GR 45110 Ioannina, Greece}

\date{}

\begin{document}

\baselineskip 18pt

\maketitle

\begin{abstract}
\absf
We present a method to solve initial and boundary value problems
using artificial neural networks. 
A trial solution of the differential equation is written
as a sum of two parts. 
The first part satisfies the initial/boundary conditions 
and contains no adjustable parameters.
The second part is constructed so as not to affect the 
initial/boundary conditions. This part involves
a feedforward neural network, containing adjustable parameters (the weights).
Hence by construction the initial/boundary conditions are satisfied and
the network is trained to satisfy the differential equation.
The applicability of this approach ranges from single ODE's,
to systems of coupled ODE's and also to PDE's. 
In this article we illustrate the method by solving a variety model problems
and present comparisons with finite elements for several cases of 
partial differential equations.
\end{abstract}

\newpage

\section{Introduction}
\textf

Many methods have been developed so far for solving differential equations.
Some of them produce a solution in the form of an array that contains 
the value of the solution at a selected group of points. Others use
basis-functions to represent the solution in analytic form and
transform the original problem usually in a system of linear equations. 
Most of the previous work in solving differential equations using neural
networks is restricted to the case of solving the linear systems of 
algebraic equations which result from the discretization of the domain.  
The solution of a linear system of equations is mapped 
onto the architecture of a Hopfield neural network. The minimization of the
network's energy function provides the solution to the system of equations
\cite{Lee90,Wan90,Yen96}.

Another approach to the solution of ordinary differential equations is
based on the fact that certain types of splines, for instance linear B-splines,
can be derived by the superposition of 
piecewise linear activation functions \cite{Mea94a,Mea94b}. 
The solution of a differential
equation using linear B-splines as basis functions, can be obtained
by solving a system of linear or non-linear equations in order  to
determine the parameters of the splines. Such a solution form is mappped 
directly on the architecture of a 
feedforward neural network by replacing each spline with the 
sum of piecewise linear activation functions that correspond to the hidden
units. 
This method considers  
local basis-functions and in general requires many splines (and consequently
network parameters) in order to yield accurate solutions. 
Furthermore it is not easy to extend these techniques to multidimensional 
domains.

In this article  we view the problem from a different angle. 
We present a general method for solving both 
ordinary differential equations (ODEs) and partial differential equations
(PDEs), 
that relies on the function approximation 
capabilities of feedforward neural networks and results in 
the construction of a solution written in a diferentiable, closed analytic
form. This form employs a feedforward neural network as the basic
approximation element, whose parameters (weights and biases) 
are adjusted to minimize an appropriate  error function.
To train the network we employ optimization techniques, which in turn 
require the computation of the gradient of the error 
with respect to the network parameters.
In the proposed approach the model function is  expressed as the sum 
of two terms: the first term satisfies the initial/boundary conditions 
and contains no adjustable parameters. 
The second term involves a feedforward neural network to be trained so as to
satisfy the differential equation. 
Since it is known that a multilayer perceptron with one hidden layer 
can approximate any function to arbitrary accuracy, it is reasonable
to consider this type of network architecture  as a candidate model for treating
differential equations. 

The employement of a neural architecture 
adds to the method many attractive features:
\begin{itemize}
\item The solution via ANN's is a {\em differentiable, closed analytic form} 
easily used in any subsequent calculation. Most other techniques offer
a discrete solution (for example predictor-corrector, or Runge-Kutta methods)
or a solution of limited differentiability (for example finite elements).
\item Such a solution is characterized by the generalization properties of
neural networks, which are known to be superior. (Comparative 
results presented in this work illustrate this point clearly.)
\item The required number of model parameters is far less than any other 
solution technique and therefore, compact solution models are obtained,
with very low demand on memory space. 
\item The method is general and can be applied to ODEs, systems of ODEs and 
to PDEs as well.
\item The method can be realized in hardware, using neuroprocessors, 
and hence offer the opportunity to tackle in real time difficult
differential equation problems arising in many engineering applications.
\item The method can also be efficiently implemented on parallel architectures.  
\end{itemize}

In the next section we describe the general formulation of the proposed
approach and derive formulas for computing the gradient  
of the error function.
Section 3 illustrates some classes of 
problems where the proposed method can be applied and 
describes the appropriate
form of the trial solution. Section 4 presents numerical examples from the
application of the technique to several test problems 
and provides details concerning the implementation of the method
and the accuracy of the obtained solution.
We also make a comparison of our results with those obtained by the finite element
method for the examined PDE problems. 
Finally, section 6 contains conclusions and directions for future research.

\section{Description of the method}

The proposed approach will be illustrated in terms of the following
general differential equation definition:
\begin{equation}
G(\vec{x},\Psi(\vec{x}),\nabla \Psi(\vec{x}),
\nabla^{2} \Psi(\vec{x}))=0, \vec{x} \in D
\end{equation}
subject to certain boundary conditions (B.Cs)
(for instance Dirichlet and/or Neumann), 
where $\vec{x}=(x_1,\ldots,x_n) \in R^n$, 
$D \subset R^n$ denotes the definition
domain and $\Psi(\vec{x})$ is the solution
to be computed. The proposed approach can be also 
applied to differential equations of higher order, but we have not
considered any problems of this kind in the present work.  
 
To obtain a solution to the above differential equation
the collocation method is adopted \cite{Kin91}
which assumes a discretization of the domain $D$ and its boundary $S$
into a set points $\hat{D}$ and $\hat{S}$ respectively. The problem is then
transformed into the following system of equations:
\begin{equation}
G(\vec{x_i},\Psi(\vec{x_i}),\nabla \Psi(\vec{x_i}),
\nabla^{2} \Psi(\vec{x_i}))=0, \forall \vec{x_i} \in \hat{D}
\end{equation}
subject to the constraints imposed by the B.Cs.

If $\Psi_t (\vec{x},\vec{p})$ denotes a trial solution 
with adjustable parameters $\vec{p}$,  the problem is transformed to: 

\begin{equation}
min_{\vec{p}}\sum_{\vec{x_i} \in \hat{D}} G(\vec{x_i},\Psi_t(\vec{x_i},\vec{p}),
\nabla \Psi(\vec{x_i},\vec{p}),
\nabla^{2} \Psi(\vec{x_i},\vec{p}))^2
\end{equation}
subject to the constraints imposed by the B.Cs.

In the proposed approach the trial solution $\Psi_t$ employs 
a feedforward neural network and the parameters  $\vec{p}$ correspond to the
weights and biases of the neural architecture. 
We choose  a form for the trial function $\Psi_t(\vec{x})$ 
such that by construction  satisfies the BCs. This is achieved by writing 
it as a sum of two terms:
\begin{equation}
\Psi_t(\vec{x})=A(\vec{x}) + F(\vec{x}, N(\vec{x},\vec{p}))
\end{equation}
where $N(\vec{x},\vec{p})$ is a single-output feedforward neural 
network with  parameters $\vec{p}$ and  $n$ input units
fed with the input vector $\vec{x}$.

The term $A(\vec{x})$ contains no adjustable parameters and
satisfies the boundary conditions. 
The second term $F$ is constructed so as not to contribute to the BCs, 
since  $\Psi_t(\vec{x})$ must also satisfy them. 
This term employs a neural network whose weights and biases
are to  be adjusted in order to deal with the minimization problem. 
Note at this point that the problem has been reduced from the original
constrained optimization problem to an unconstrained one
(which is much easier to handle) due to the choice of the form of
the trial solution that satisfies by construction the B.Cs.

In the next section we present
a systematic way to construct  the trial solution,
i.e. the functional forms of both $A$ and $F$. 
We treat several common cases that one frequently encounters in 
various scientific fields.
As indicated by our experiments, the approach based on the above 
formulation is  very effective and provides in reasonable computing time
accurate solutions with impressive generalization (interpolation) properties.

\subsection{Gradient Computation}

The efficient minimization of equation (3) 
can be considered as a procedure of training the neural network where the error 
corresponding to each input vector $\vec{x_i}$ is the value $G(\vec{x_i})$
which has to become zero. Computation of this
error value involves not only the network output (as is the case in conventional
training) but also the derivatives of the output with respect to any of its
inputs. Therefore, in computing the gradient of the error with respect to the
network weights, we need to compute not only the gradient of the network
but also the gradient of the network derivatives with respect to its inputs.

Consider a multilayer perceptron with $n$ input units, 
one hidden layer with $H$ sigmoid units and a linear output unit. 
The  extension to the case of more than one hidden layers can be 
obtained accordingly.
For a given input vector $\vec{x}=(x_1,
\ldots, x_n)$ the output of the network is $N=\sum_{i=1}^H v_i
\sigma(z_i)$ where $z_i= \sum_{j=1}^N w_{ij}x_j+u_i$, $w_{ij}$ denotes 
the weight from the input unit $j$ to the
hidden unit $i$, 
$v_i$ denotes the weight from the hidden unit $i$ to the output,
$u_i$ denotes the bias of hidden unit $i$ and $\sigma(z)$ is the 
sigmoid transfer function. It is straightforward to show that:
\begin{equation}
\frac{\partial^k N}{\partial x^{k}_j}=\sum_{i=1}^{H} v_iw_{ij}^k \sigma_i^{(k)}
\end{equation}
where $\sigma_i=\sigma(z_i)$ and $\sigma^{(k)}$ denotes
the $k^{th}$ order derivative of the sigmoid.
Moreover it is readily verifiable that:
\begin{equation}
\frac{\partial^{\lambda_1}}{\partial x_1^{\lambda_1}}
\frac{\partial^{\lambda_2}}{\partial x_2^{\lambda_2}}
\ldots
\frac{\partial^{\lambda_n}}{\partial x_2^{\lambda_n}}N=\sum_{i=1}^n v_iP_i
\sigma_i^{(\Lambda)}
\end{equation}
where 
\begin{equation}
P_i=\prod_{k=1}^n w_{ik}^{\lambda_k}
\end{equation}
and $\Lambda=\sum_{i=1}^n \lambda_i$.

Equation (6) indicates that the derivative of the network
with respect to any of its inputs is equivalent to  a feedforward 
neural network $N_g(\vec{x})$ with one hidden layer, having 
the same values for the weights $w_{ij}$ and thresholds $u_i$ and 
with each weight $v_i$ being replaced with $v_iP_i$. 
Moreover the transfer function of each hidden
unit is replaced with the $\Lambda^{th}$ order derivative of the sigmoid.

Therefore the gradient of $N_g$ with respect
to the parameters of the original network can be easily obtained as:
\begin{equation}
\frac{\partial N_g}{\partial v_i}=P_i\sigma_i^{(\Lambda)}
\end{equation}
\begin{equation}
\frac{\partial N_g}{\partial u_i}=v_iP_i\sigma_i^{(\Lambda+1)}
\end{equation} 
\begin{equation}
\frac{\partial N_g}{\partial w_{ij}}=x_jv_iP_i\sigma_i^{(\Lambda+1)}
+ v_i \lambda_j w_{ij}^{\lambda_j - 1}
(\prod_{k=1,k\neq j}w_{ik}^{\lambda_k})\sigma_i^{(\Lambda)}
\end{equation} 

Once the derivative of the error with respect
to the network parameters has been defined
it is then straightforward to employ almost any minimization technique.
For example it is possible to use either the steepest  
descent (i.e. the backpropagation algorithm or any of its variants), or 
the conjugate gradient method or other techniques proposed in the literature.
In our experiments we have employed the BFGS method \cite{BFGS} that is 
quadraticly convergent and has demonstrated excellent 
performance. It must also be noted that for a given grid point 
the derivatives of each network (or gradient network) with respect 
to the parameters  may be obtained simultaneously in the case 
where parallel harware is available. 
Moreover, in the case of backpropagation, the on-line or batch
mode of weight updates may be employed.

\section{Illustration of the method}

\subsection{Solution of single ODEs and Systems of coupled ODEs}

To illustrate the method, we consider the {\em first order ODE}:
$$ \frac{d\Psi(x)}{dx} = f(x,\Psi) $$
with $x \in [0,1]$ and with the IC  $\Psi(0)=A$.

A trial solution is written as:
\begin{equation}
\Psi_t(x)=A + x N(x,\vp)
\end{equation}
where $  N(x,\vp)$ is the output of a feedforward
neural network with one input unit for $x$ 
and weights $\vp$. 
Note that $\Psi_t(x)$ satisfies the IC by construction. 
The error quantity to be minimized is given by:
\begin{equation}
E[\vp]=\sum_i\{ \frac{d\Psi_t(x_i)}{dx}-f(x_i,\Psi_t(x_i))\}^2
\end{equation}
where the $x_i$'s are points in  $[0,1]$.
Since $d \Psi_t(x)/dx = N(x, \vec{p}) + x dN(x,\vec{p})/dx$, it is
straightforward to compute the gradient of the error with respect to
the parameters $\vec{p}$ using equations (5)-(10). The same holds for
all subsequent model problems.  

The same procedure can be applied to the {\em second order ODE}:
$$ \frac{d^2\Psi(x)}{dx^2} = f(x,\Psi,\frac{d\Psi}{dx}) $$
For the {\em initial value} problem:  $\Psi(0)=A$ and 
$ \frac{d}{dx}\Psi(0)=A' $, the trial solution can be cast as:
\begin{equation}
\Psi_t(x)= A + A'x + x^2N(x,\vp) 
\end{equation}
\noindent For the {\em two point Dirichlet} BC: $\Psi(0)=A$ and $\Psi(1)=B$,
the trial solution is written as:
\begin{equation}
\Psi_t(x)= A(1-x) + B x + x(1-x)N(x,\vp)
\end{equation}
In the above two cases of second order ODEs the error function to be minimized
is given by equation (12).

For {\em systems of $K$ first order ODEs} 
$$ \frac{d\Psi_i}{dx} = f_i(x,\Psi_1,\Psi_2,\ldots \Psi_K)$$
with $\Psi_i(0) = A_i$, $(i=1, \ldots, K)$ we consider one neural network
for each trial solution $\Psi_{t_i}$ $(i=1, \ldots, K)$ which 
is written as:
\begin{equation}
\Psi_{t_i}(x)=A_i + x N_i(x,\vp_i)
\end{equation}
and we minimize the following error quantity:
\begin{equation}
E[\vp]=\sum_{k=1}^K\sum_i\{ \frac{d\Psi_{t_k}(x_i)}{dx}
   -f_k(x_i,\Psi_{t_1},\Psi_{t_2}, \ldots, \Psi_{t_K})\}^2 
\end{equation}
   
\subsection{Solution of single PDEs}

We treat here two--dimensional problems only. However
it is straightforward to extend the method to more dimensions.
For example consider the {\em Poisson equation}:
$$ \frac{\partial^2}{\partial x^2}\Psi(x,y) +
\frac{\partial^2}{\partial y^2}\Psi(x,y) = f(x,y) $$
$x \in [0,1], y \in [0,1]$  
with {\em Dirichlet} BC: $\Psi(0,y)=f_0(y)$, $ \Psi(1,y)=f_1(y)$ and 
$\Psi(x,0)=g_0(x)$, $\Psi(x,1)=g_1(x)$. The trial solution is written as:
\begin{equation}
\Psi_t(x,y)=A(x,y)+x(1-x)y(1-y)N(x,y,\vp)
\end{equation} 
where $A(x,y)$ is chosen so as to satisfy the BC, namely:
\begin{equation}
A(x,y)=(1-x)f_0(y)+xf_1(y)+(1-y)\{g_0(x)-[(1-x)g_0(0)+xg_0(1)]\}
+ y\{g_1(x)-[(1-x)g_1(0)+xg_1(1)]\} 
\end{equation}

For {\em mixed boundary conditions} of the form:
$\Psi(0,y)=f_0(y)$, $ \Psi(1,y)=f_1(y)$, $\Psi(x,0)=g_0(x)$
 and $\frac{\partial}{\partial y}\Psi(x,1)=g_1(x)$
(i.e. {\em Dirichlet} on part of the boundary and {\em Neumann} elsewhere), the 
trial solution is written as:
\begin{equation}
\Psi_t(x,y)=B(x,y)+x(1-x)y[N(x,y,\vp)-N(x,1,\vp)-
\frac{\partial}{\partial y}N(x,1,\vp)] 
\end{equation}
and $B(x,y)$ is again chosen so as to satisfy the BCs:
\begin{equation}
B(x,y)=(1-x)f_0(y)+xf_1(y)+g_0(x)-[(1-x)g_0(0)+xg_0(1)]+ 
y\{g_1(x)-[(1-x)g_1(0)+xg_1(1)]\} 
\end{equation}

Note that the second term of the trial solution does not affect
the boundary conditions since it vanishes at the part of the boundary
where Dirichlet BCs are imposed and its  gradient component normal
to the boundary vanishes at the part of the boundary where Neumann BCs
are imposed.

In all the above PDE problems the error to be minimized is given by:
\begin{equation}
E[\vec{p}]=\sum_i  \{ \frac{\partial^2}{\partial x^2}\Psi(x_i,y_i) +
\frac{\partial^2}{\partial y^2}\Psi(x_i,y_i) - f(x_i,y_i)\}^2 
\end{equation}
where $(x_i,y_i)$ are points in $[0,1]\times [0,1]$.

\section{Examples}
In this section we report on the solution of a number of model problems.
In all cases we used a multilayer perceptron having one hidden layer
with 10 hidden units and one linear output unit. The sigmoid activation of each
hidden unit is $\sigma(x)= \frac{1}{1+e^{-x}}$.
For each test problem the exact analytic solution $\Psi_a(\vec{x})$
was known in advance. Therefore we test the accuracy of the obtained
solutions by computing the deviation $\Delta \Psi(\vec{x})=
\Psi_t(\vec{x})-\Psi_a(\vec{x})$.
To perform the error minimization we employed the {\em Merlin} 
\cite{Merlin1,Merlin2} optimization package. {\em Merlin} provides an 
environment for multidimensional continuous function optimization.
From the several algorithms that are implemented therein, the 
Quasi--Newton {\em BFGS} \cite{BFGS} method seemed to perform better
in these kind of problems and hence we used it in all of our experiments.
A simple criterion for the gradient norm was used for termination.

In order to illustrate the 
characteristics of the solutions provided by the neural
method, we provide figures displaying the corresponding deviation
$\Delta \Psi(\vec{x})$ 
both at the few points (training points)
that were used for training and at many other points (test points) of
the domain of each equation. The second kind of figures are of major importance
since they show the interpolation capabilities of the neural 
solutions which seem
to be superior compared 
to other solutions. Moreover, in the case of ODEs we also
consider points outside the training interval in order to obtain an estimate
of the extrapolation performance of the obtained solution. 

\subsection{ODEs and systems of ODEs}

\subsubsection{\bf Problem 1}

$$ \frac{d}{dx}\Psi + (x+\frac{1+3x^2}{1+x+x^3})\Psi = 
x^3+2x+x^2\frac{1+3x^2}{1+x+x^3} $$
with $\Psi(0)=1$ and $x \in [0,1]$. The analytic solution is 
$\Psi_a(x)=\frac{e^{-x^2/2}}{1+x+x^3} +x^2$ and is displayed in Figure 1a.
According to equation (11) the trial neural form of the solution
is taken to be: $\Psi_t(x)=1+xN(x,\vp)$. The network was trained 
using a grid of 10 equidistant points in [0,1].
Figure 2 displays the deviation $\Delta \Psi(x)$ from the exact solution 
corresponding at the grid points
(small circles) and the deviation at many other points in $[0,1]$ as well as
outside that interval (dashed line). It is clear that
the solution is of high accuracy, although training was performed using a
small number of points. Moreover, the extrapolation error remains low for points
near the equation domain. 
   
\subsubsection{\bf Problem 2}
$$\frac{d}{dx}\Psi + \frac{1}{5}\Psi = e^{-\frac{x}{5}}cos(x)$$
with $\Psi(0)=0$ and $x \in [0,2]$. The analytic solution is
$\Psi_a (x)=e^{-\frac{x}{5}}sin(x)$ and is presented in Figure 1b.
The trial neural form is:
$\Psi_t(x)=xN(x,\vp)$ according to equation (11). 
As before we used a 
grid of 10 equidistant points in [0,2] to perform the training.
In analogy with the previous case, 
Figure 3 display the deviation $\Delta \Psi(x)$ at the grid points
(small circles) and at many other points 
inside and outside the training interval
(dashed line). 

\subsubsection{\bf Problem 3}
$$ \frac{d^2}{dx^2}\Psi + \frac{1}{5}\frac{d}{dx}\Psi + \Psi = -\frac{1}{5}e^{-\frac{x}{5}}cosx$$
Consider the {\em initial value} problem: 
$\Psi(0)=0$ and $\frac{d}{dx}\Psi(0)=1$ with $x \in [0,2]$.
The exact solution is: $\Psi(x)=e^{-\frac{x}{5}}sin(x)$ and the trial neural form is:
$\Psi_t(x)=x+x^2N(x,\vp)$ (from equation (13)). \\
Consider also the {\em boundary value} problem: 
$\Psi(0)=0$ and $\Psi(1)=sin(1)e^{-\frac{1}{5}}$, $x \in [0,1]$.
The exact solution is the same as above, but the appropriate trial neural form is:
$\Psi_t(x)=x sin(1)e^{-\frac{1}{5}}+x(1-x)N(x,\vp)$ (from equation (14)).

Again as before we used a grid of 10 equidistant points and the plots
of the deviation from the exact 
solution  are displayed at Figures 4 and 5 for the initial value
and boundary value problem respectively. 
The interpretation of the figures is the
same as in the previous cases.

From all the above cases it 
is clear that method can handle effectively all kinds
of ODEs and provides analytic solutions that remain to be of the same accuracy at
points other from the training ones.

\subsubsection{\bf Problem 4}
Consider the  system of two coupled first order ODEs:
$$  \frac{d}{dx}\Psi_1 = cos(x) + \Psi_1^2+\Psi_2 -(1+x^2 + sin^2(x))$$
$$  \frac{d}{dx}\Psi_2 = 2x-(1+x^2)sin(x) + \Psi_1\Psi_2 $$
with $x \in [0,3]$ and $\Psi_1(0)=0$ and  $\Psi_2(0)=1$. The analytic solutions are
$\Psi_{a1}(x)=sin(x)$ and $\Psi_{a2}(x)= 1+x^2$ and are displayed at
Figure 6a and 6b, respectively. Following 
equation (15) the trial neural solutions are:
$\Psi_{t_1}(x)=xN_1(x,\vp_1)$ and $\Psi_{t_2}(x)=1+xN_2(x,\vp_2)$
where the networks $N_1$ and $N_2$ have the same architecture as
in the previous cases. Results concerning the accuracy of the obtained solutions
at the grid points (small circles) and at many other points (dashed line) are
presented in Figure 7.   

\subsection{PDEs}

We consider boundary value problems with Dirichlet and Neumann BCs.
All subsequent problems were defined on the domain $[0,1]\times [0,1]$ and
in order to perform training we consider a mesh of 100 points obtained 
by considering 10 equidistant points of the domain $[0,1]$ of each variable.
In analogy with the previous cases the neural architecture 
was considered to be a MLP
with two inputs (accepting the coordinates $x$ and $y$ of each point), 
10 sigmoid hidden units and one linear output unit.
 
\subsubsection{\bf Problem 5}
$$ \nabla^2\Psi(x,y) = e^{-x}(x-2+y^3+6y)$$
with $x,y \in [0,1]$ and the Dirichlet BCs:
$\Psi(0,y)=y^3$, $\Psi(1,y)=(1+y^3)e^{-1}$ and
$\Psi(x,0)=xe^{-x}$, $\Psi(x,1)=e^{-x}(x+1)$.
The analytic solution is $\Psi_a (x,y)=e^{-x}(x+y^3)$ and is displayed in
Figure 8.
Using equation (17) the trial neural form must be written:
$\Psi_t(x,y)= A(x,y) + x(1-x)y(1-y)N(x,y,\vp) $ and $A(x,y)$ is obtained by 
direct substitution in the general form given by equation (18):
$$ A(x,y)=(1-x)y^3+x(1+y^3)e^{-1}+(1-y)x(e^{-x}-e^{-1})+y[(1+x)e^{-x}-(1-x-2xe^{-1})]$$
Figure 9 presents the deviation $\Delta \Psi(x,y)$ 
of the obtained solution at the 100
grid points that were selected for training while Figure 10 displays the deviation 
at 900 other points of the equation domain. It clear that the solution is very
accurate and the accuracy remains high at all points of the domain.

\subsubsection{\bf Problem 6}
$$\nabla^2\Psi(x,y) = e^{-\frac{ax+y}{5}}\{[-\frac{4}{5}a^3x-\frac{2}{5}+2a^2]
cos(a^2x^2+y) + [\frac{1}{25}-1 -4a^4x^2+\frac{a^2}{25} ]sin(a^2x^2+y)\}                  
$$
with $a=3$, $x,y \in [0,1]$ 
and the Dirichlet BCs as defined by the exact solution
$\Psi_a(x,y)=e^{-\frac{ax+y}{5}}sin(a^2x^2+y)$ (presented in Figure 11).
Again the trial neural form is:
$\Psi_t(x,y)= A(x,y) + x(1-x)y(1-y)N(x,y,\vp) $ and $A(x,y)$ is obtained
similarly by direct substitution in equation (18).
Accuracy results are presented in Figure 12 for the training points and
in Figure 13 for test points. It can be shown that the accuracy is not
the same as in the previous example, but it can be improved further by considering
a neural network with more than 10 hidden units. From the figures it is also
clear that the test error lies in the same range as the training error.
 
\subsubsection{\bf Problem 7}
$$ \nabla^2\Psi(x,y) =(2-\pi ^2 y^2)sin(\pi x)$$
with $x,y \in [0,1]$ and with mixed BCs:
$\Psi(0,y)=0$, $\Psi(1,y)=0$ and
$\Psi(x,0)=0$, $\frac{\partial}{\partial y}\Psi(x,1)=2sin(\pi x)$.
The analytic solution is 
$\Psi_a(x,y)=y^2 sin(\pi x)$ and is presented in Figure 14. 
The trial neural form is specified according to
equation (19)
$$\Psi_t(x,y)= B(x,y) + x(1-x)y[N(x,y,\vp)-N(x,1,\vp)-
\frac{\partial}{\partial y} N(x,1,\vp)]$$
where $B(x,y)$ is obtained by direct substitution in equation (20).
The accuracy of the neural solution is depicted in Figures 
15 and 16 for training
and test points respectively.
 
\subsubsection{\bf Problem 8}
This is an example of a {\em non-linear} PDE.
$$ \nabla^2\Psi(x,y) + \Psi(x,y)\frac{\partial}{\partial y}\Psi(x,y) =
   sin(\pi x)(2-\pi ^2 y^2 + 2y^3 sin(\pi x)) $$ 
with the same mixed BCs as in the previous problem. The exact solution is again 
$\Psi_a(x,y)=y^2 sin(\pi x)$ and the parametrization of the trial neural form is 
the same as in problem 7. No plots of  the accuracy are presented since they 
are almost the same with those of problem 7. 

\begin{figure}
\centerline{\epsfysize=8cm\epsfxsize=13cm\epsffile{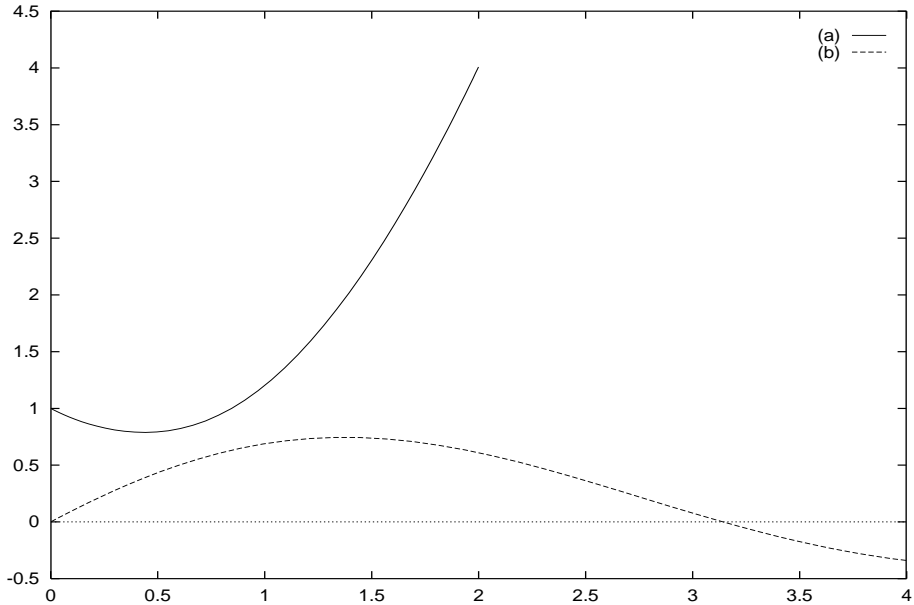}}
\caption{Exact solutions of ODE problems}
\end{figure}

\begin{figure}
\centerline{\epsfysize=8cm\epsfxsize=13cm\epsffile{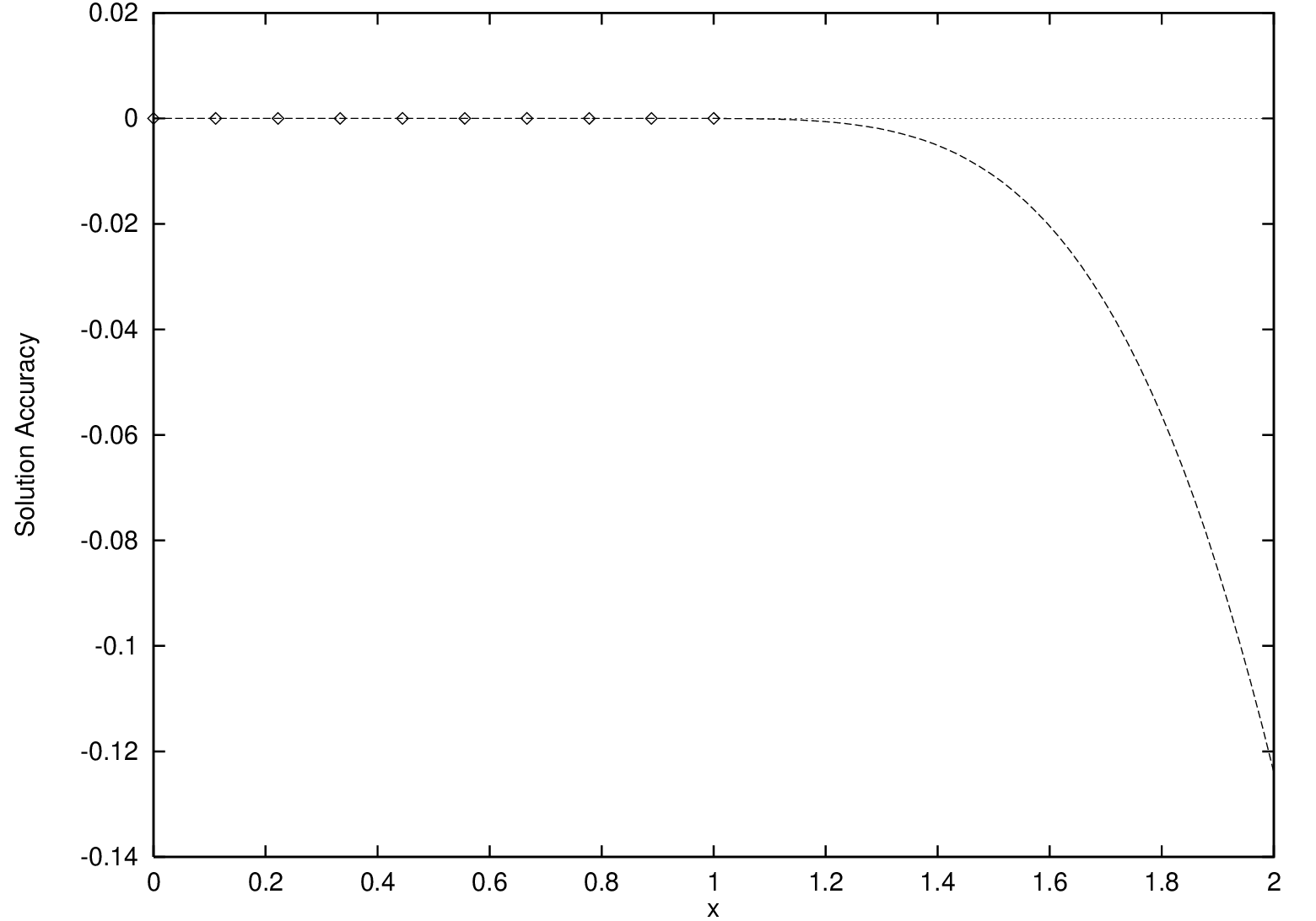}}
\caption{Problem 1: Accuracy of the computed solution.}
\end{figure}

\begin{figure}
\centerline{\epsfysize=8cm\epsfxsize=13cm\epsffile{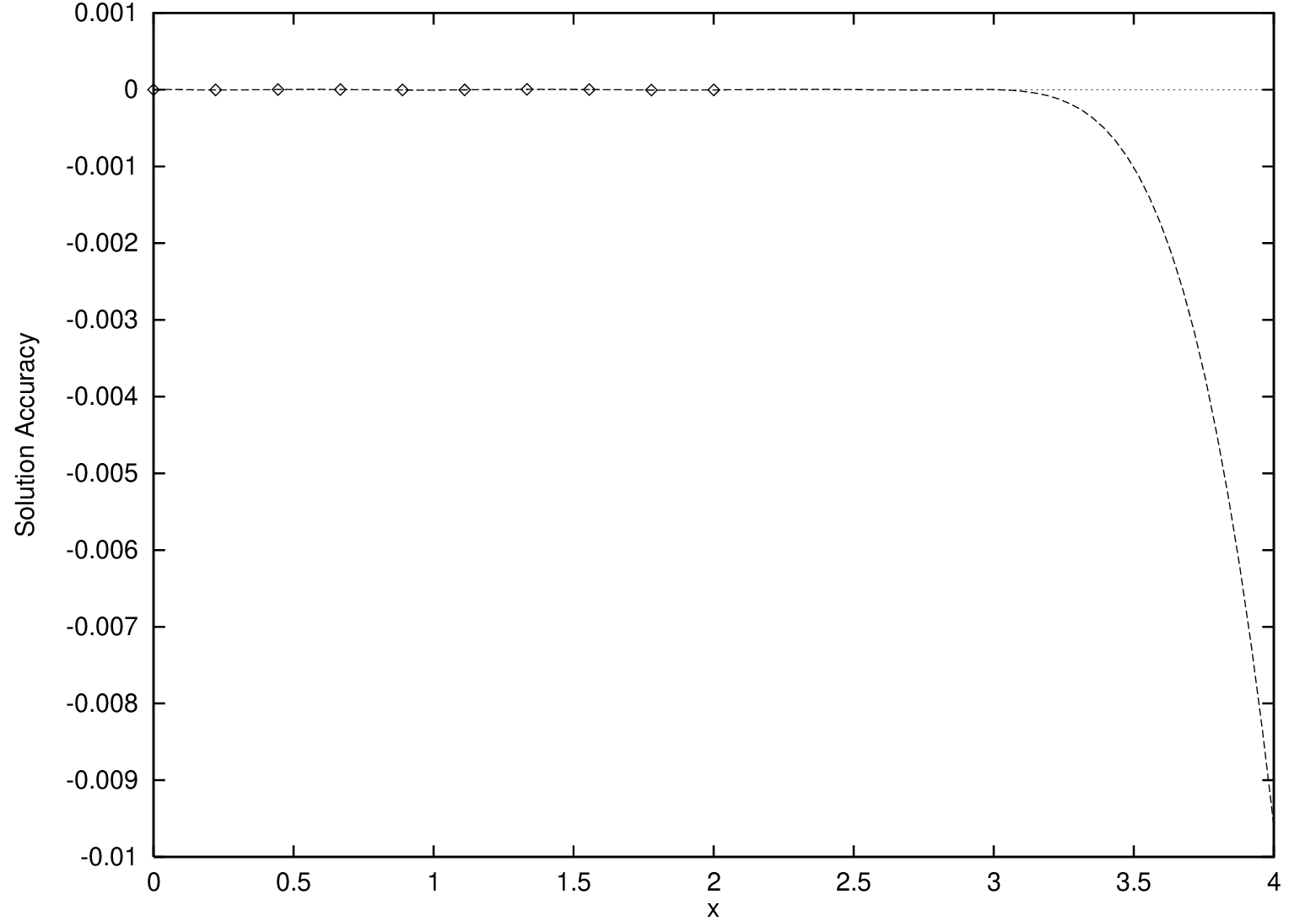}}
\caption{Problem 2: Accuracy of the computed solution.}
\end{figure}

\begin{figure}
\centerline{\epsfysize=8cm\epsfxsize=13cm\epsffile{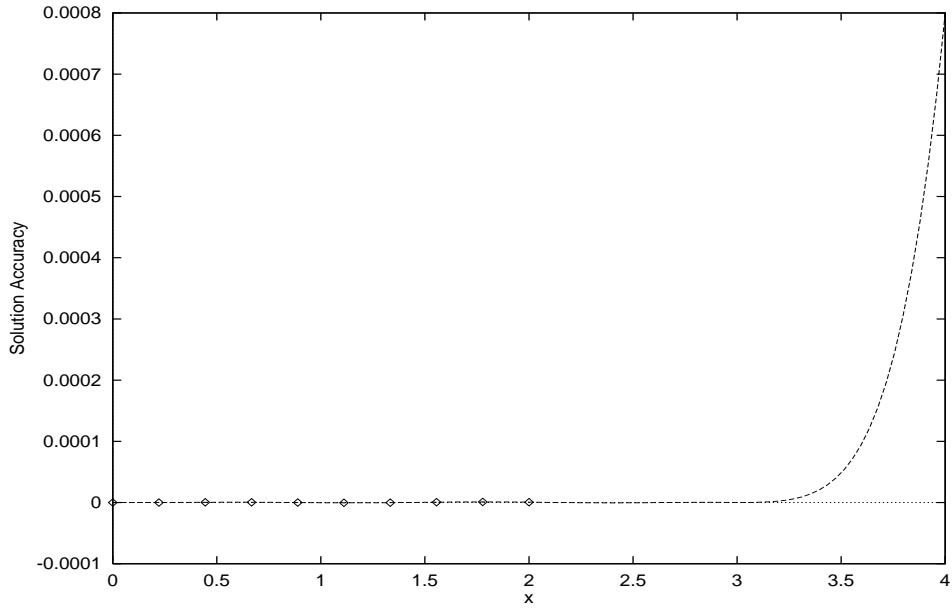}}
\caption{Problem 3 with initial conditions: Accuracy of the computed solution.}
\end{figure}

\begin{figure}
\centerline{\epsfysize=8cm\epsfxsize=13cm\epsffile{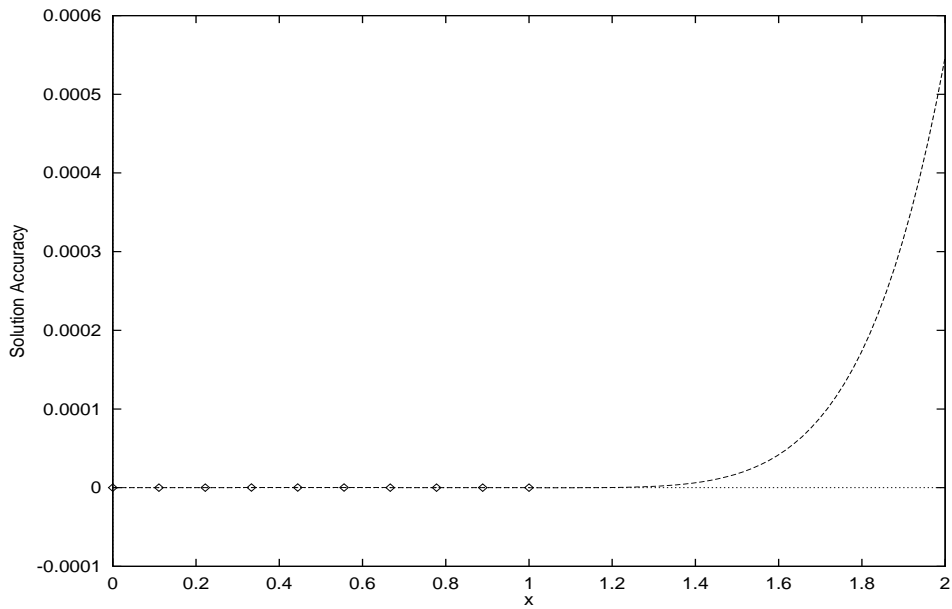}}
\caption{Problem 3 with boundary conditions: Accuracy of the computed solution.}
\end{figure}

\begin{figure}
\centerline{\epsfysize=8cm\epsfxsize=13cm\epsffile{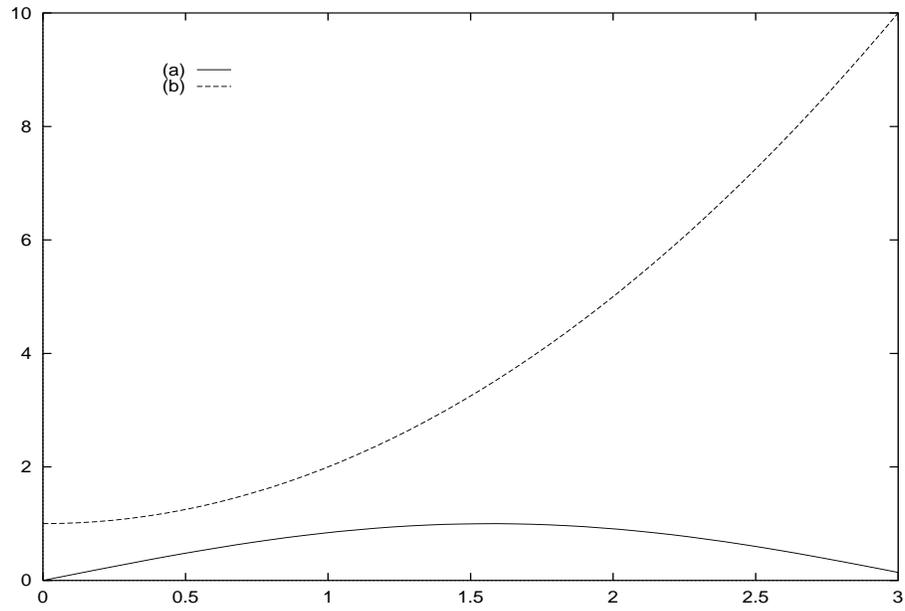}}
\caption{Exact solutions of the system of coupled ODEs.}
\end{figure}

\begin{figure}
\centerline{\epsfysize=8cm\epsfxsize=13cm\epsffile{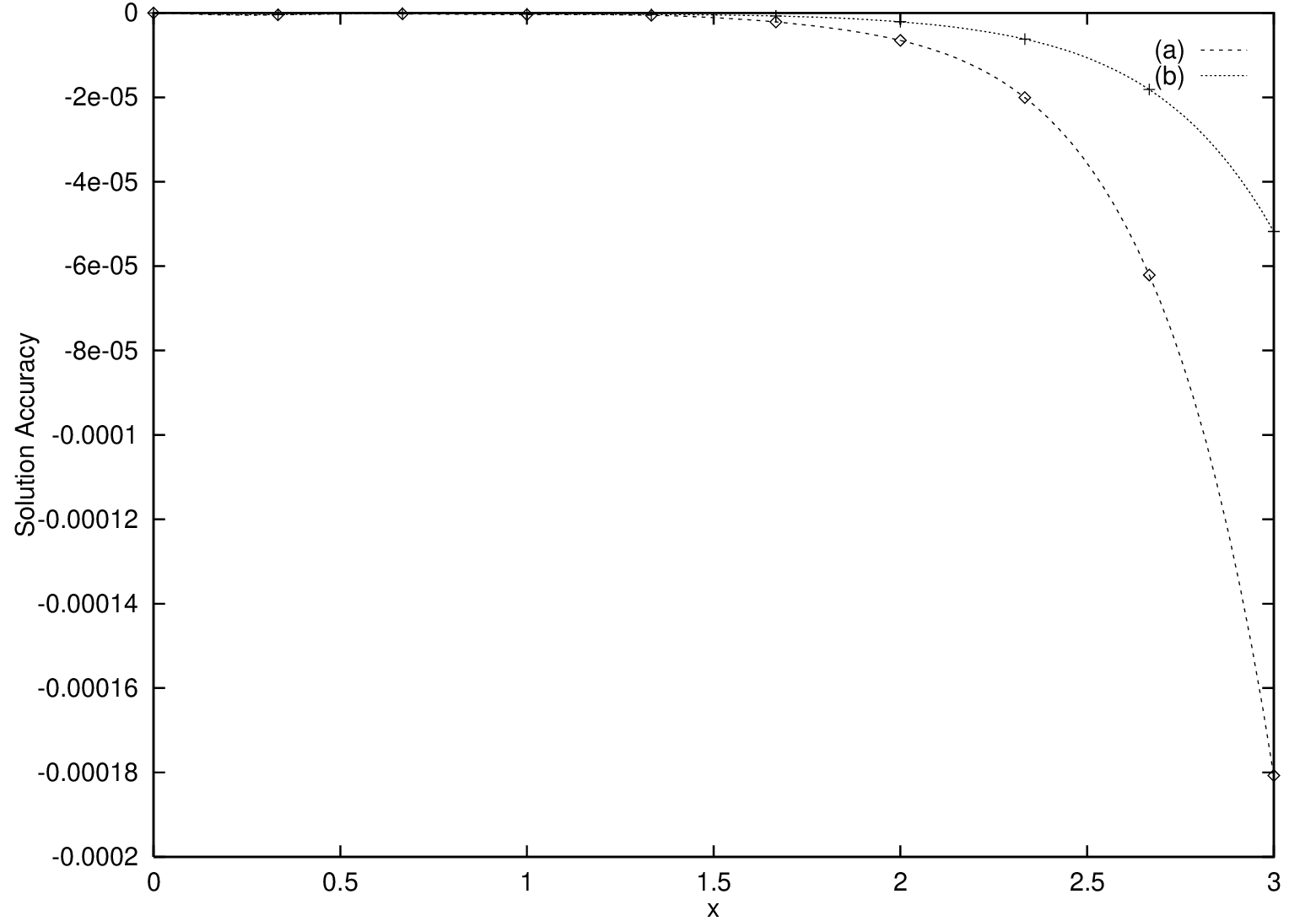}}
\caption{Problem 4: Accuracy of the computed solutions.}
\end{figure}

\begin{figure}
\centerline{\epsfysize=8cm\epsfxsize=13cm\epsffile{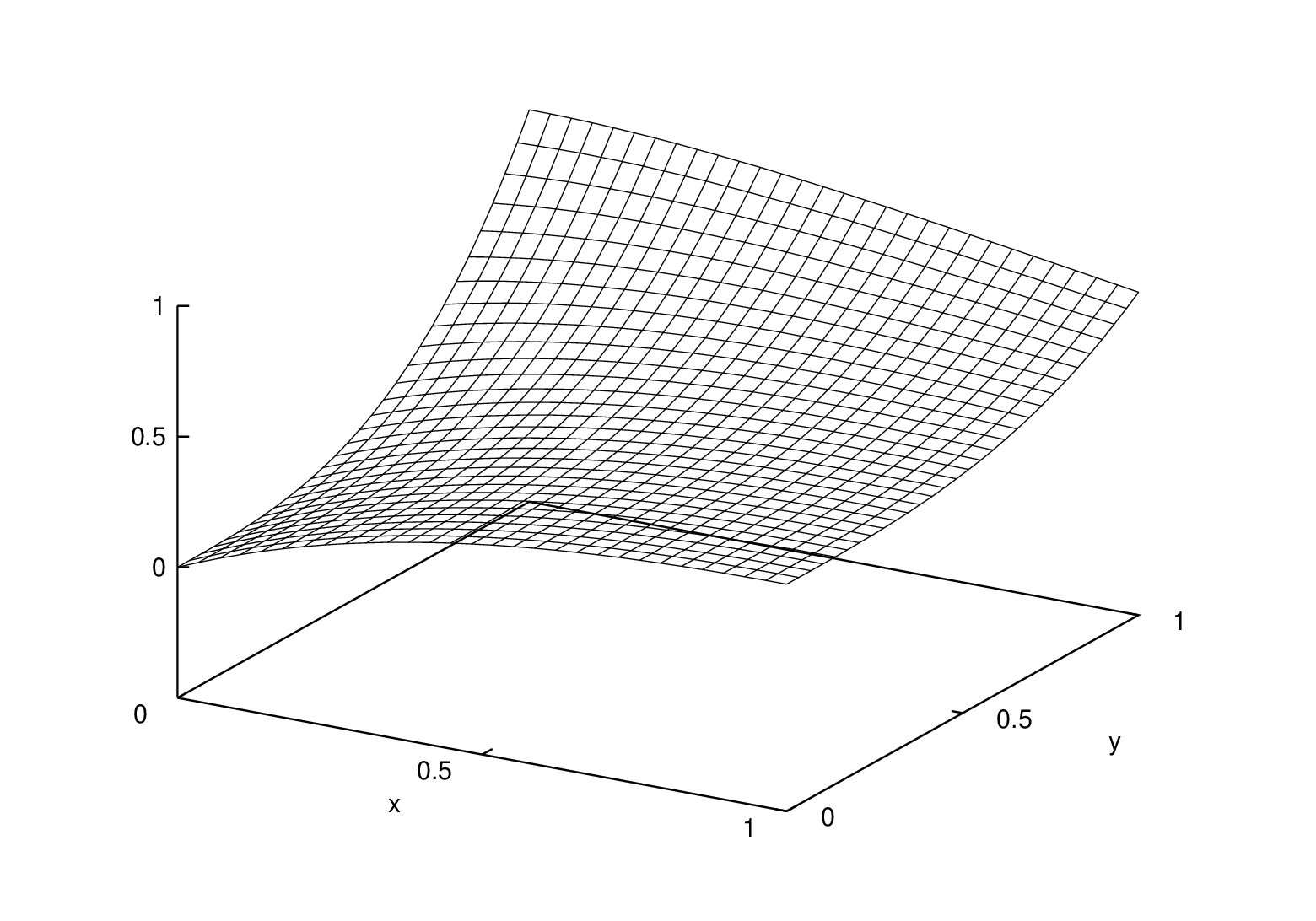}}
\caption{Exact solution of PDE problem 5.}
\end{figure}

\begin{figure}
\centerline{\epsfysize=8cm\epsfxsize=13cm\epsffile{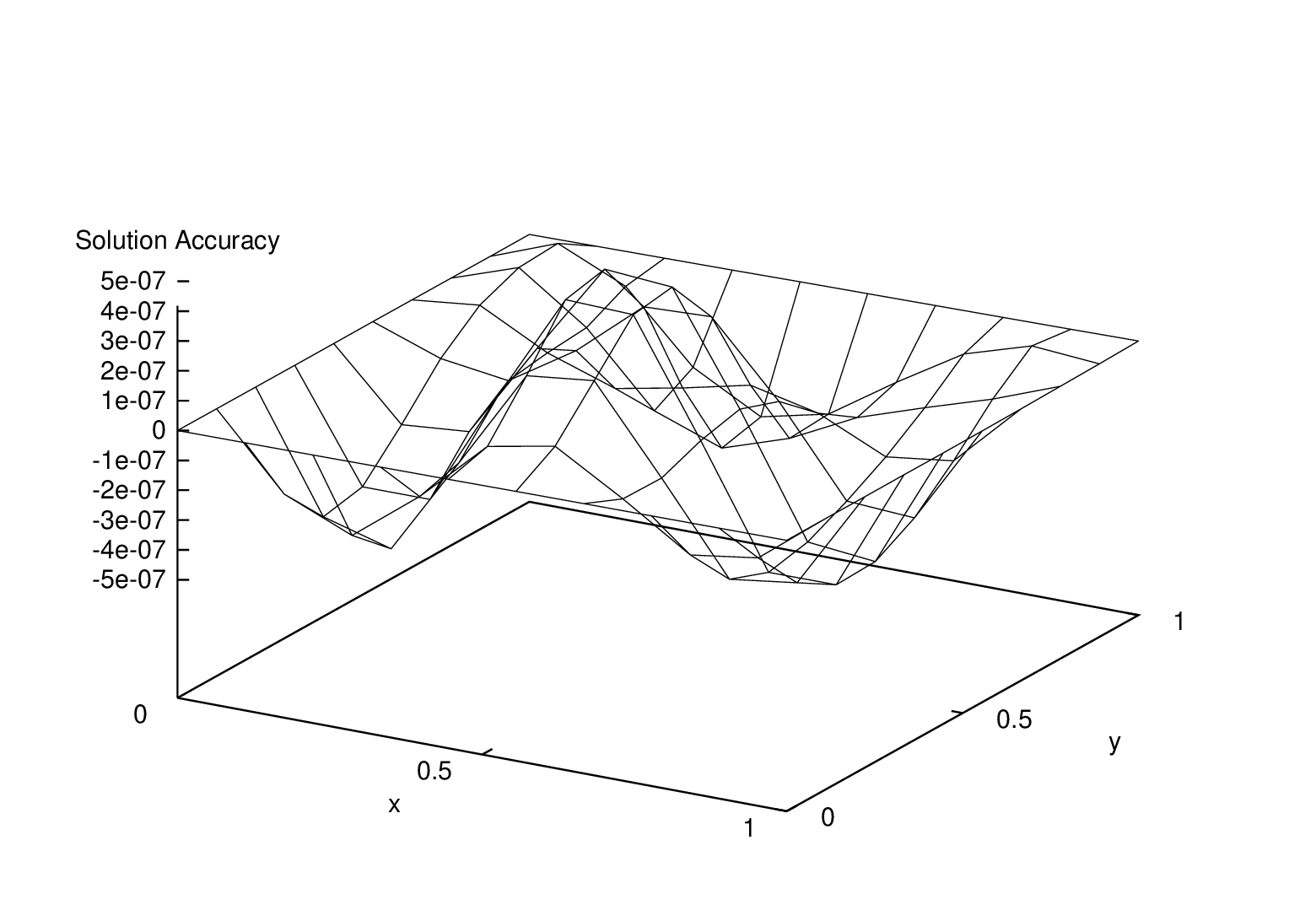}}
\caption{Problem 5: Accuracy of the computed solution at the training
points.}
\end{figure}

\begin{figure}
\centerline{\epsfysize=8cm\epsfxsize=13cm\epsffile{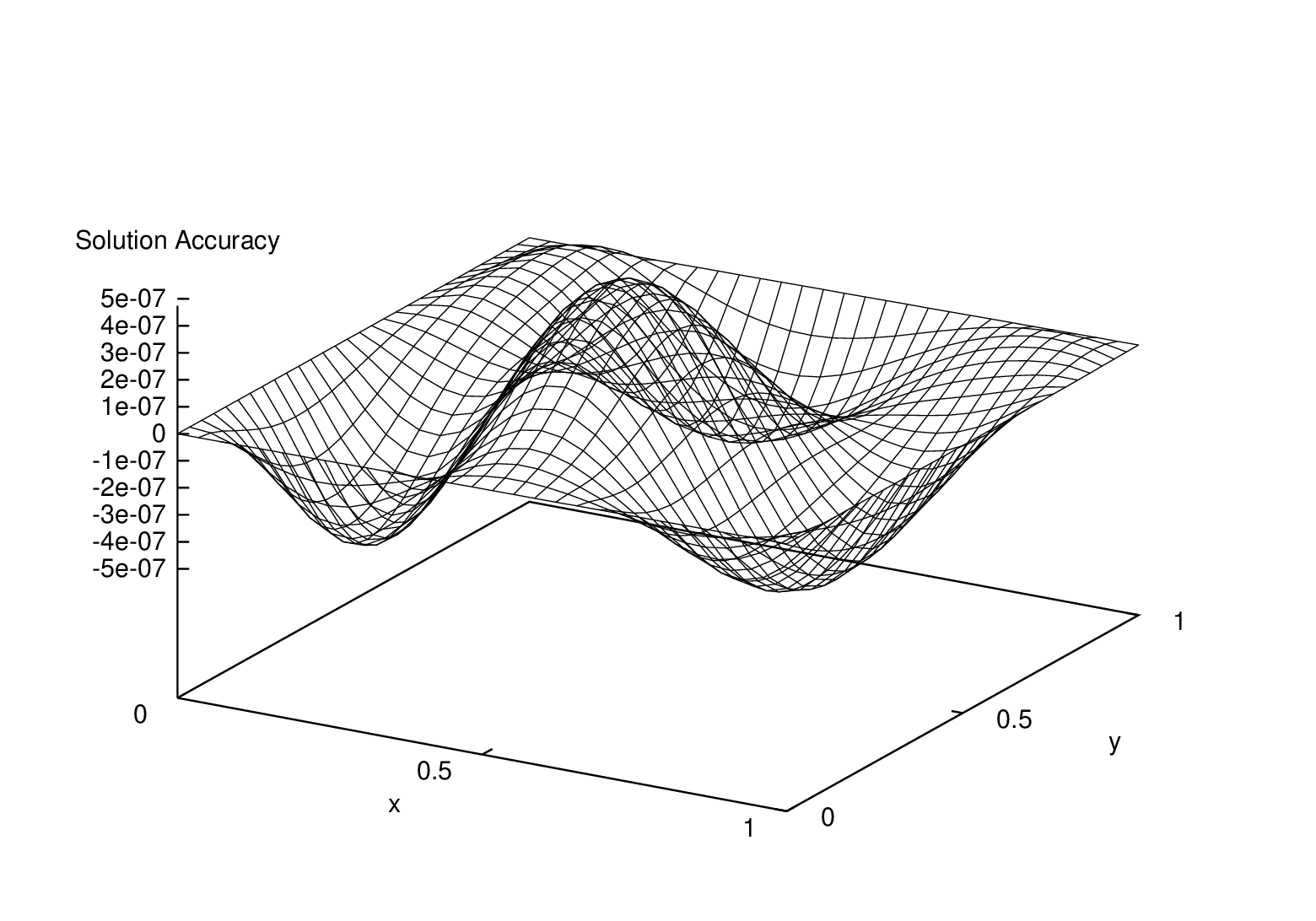}}
\caption{Problem 5: Accuracy of the computed solution at the test
points.}
\end{figure}

\begin{figure}
\centerline{\epsfysize=8cm\epsfxsize=13cm\epsffile{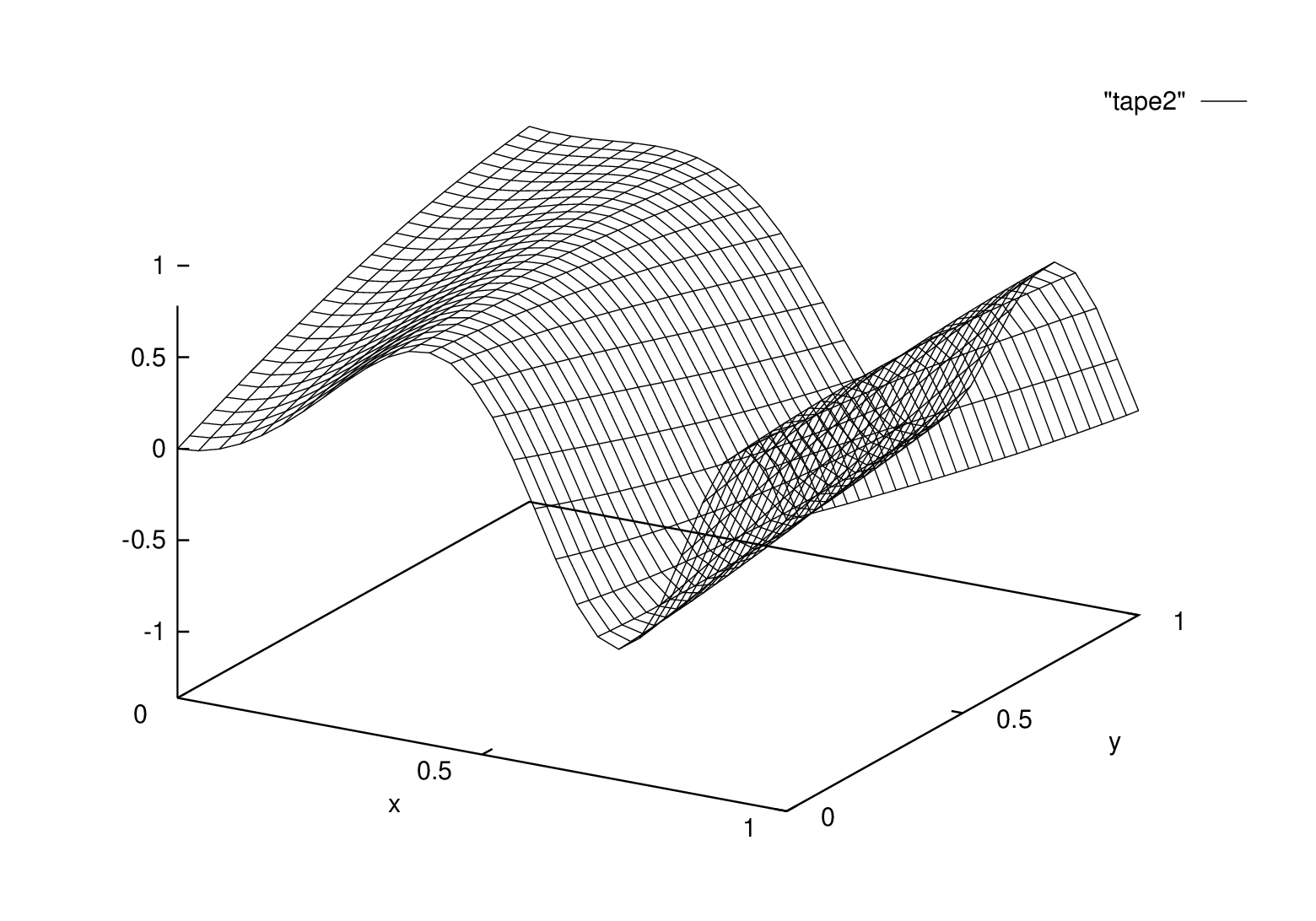}}
\caption{Exact solution of PDE problem 6.}
\end{figure}

\begin{figure}
\centerline{\epsfysize=8cm\epsfxsize=13cm\epsffile{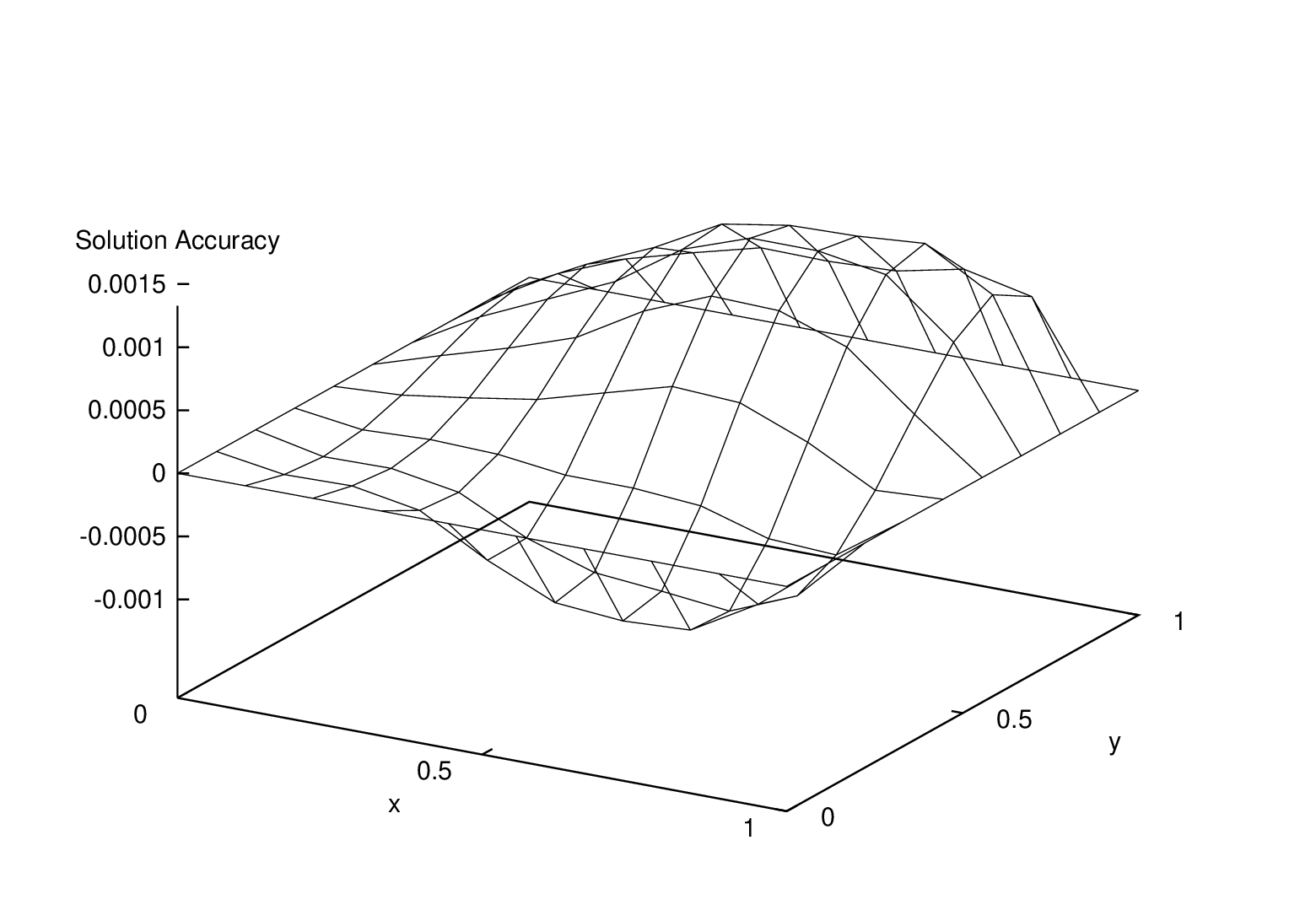}}
\caption{Problem 6: Accuracy of the computed solution at the training points.}
\end{figure}

\begin{figure}
\centerline{\epsfysize=8cm\epsfxsize=13cm\epsffile{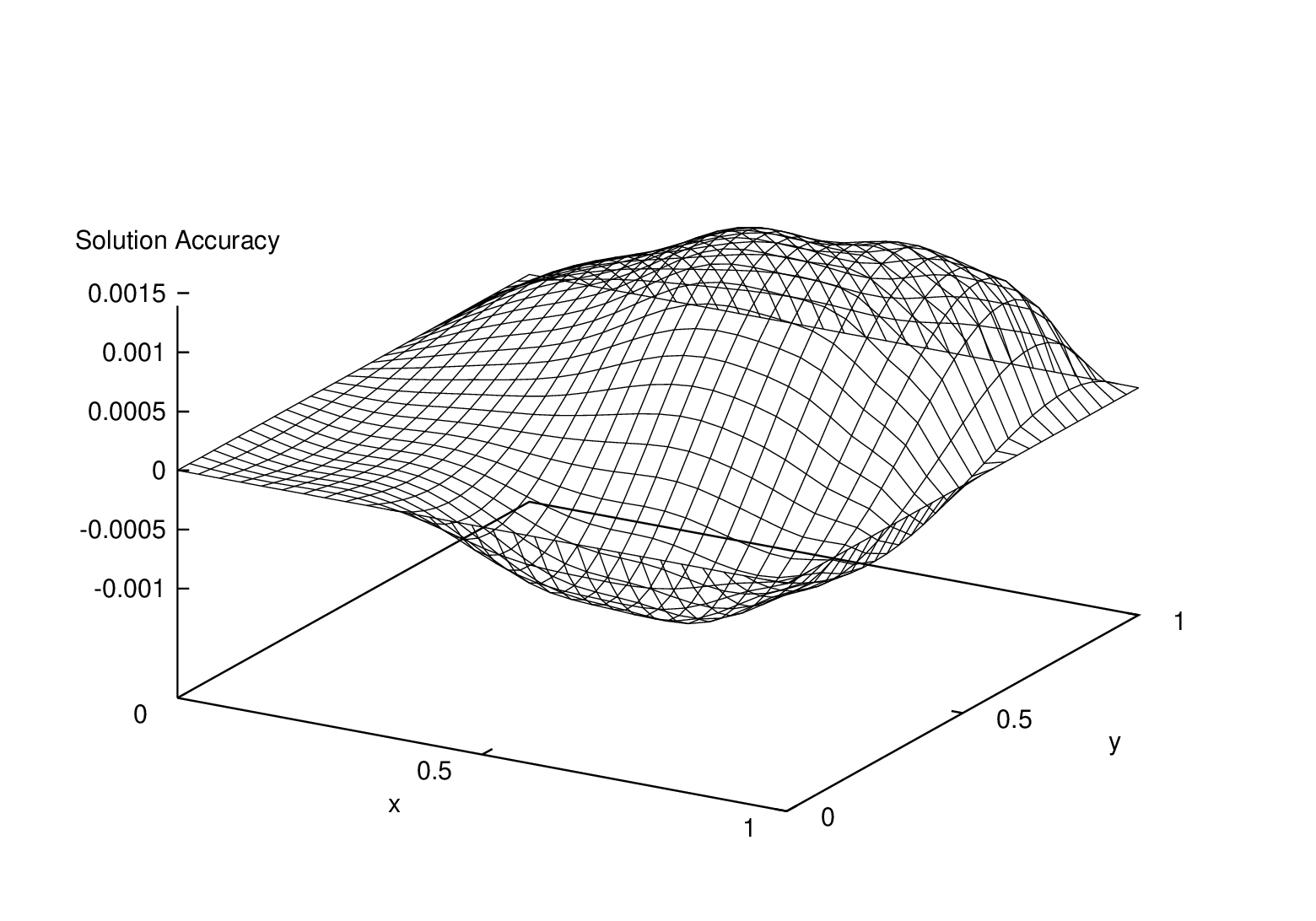}}
\caption{Problem 6: Accuracy of the computed solution at the test points.}
\end{figure}

\begin{figure}
\centerline{\epsfysize=8cm\epsfxsize=13cm\epsffile{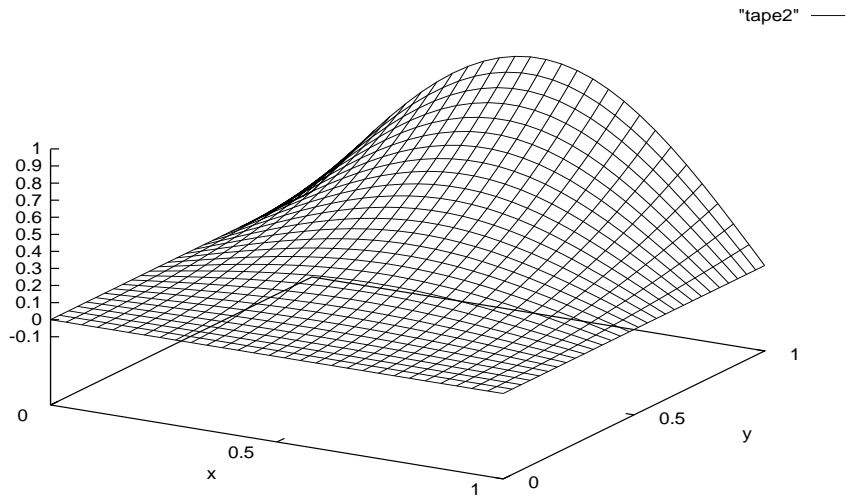}}
\caption{Exact solution of PDE problems 7 and 8.}
\end{figure}

\begin{figure}
\centerline{\epsfysize=8cm\epsfxsize=13cm\epsffile{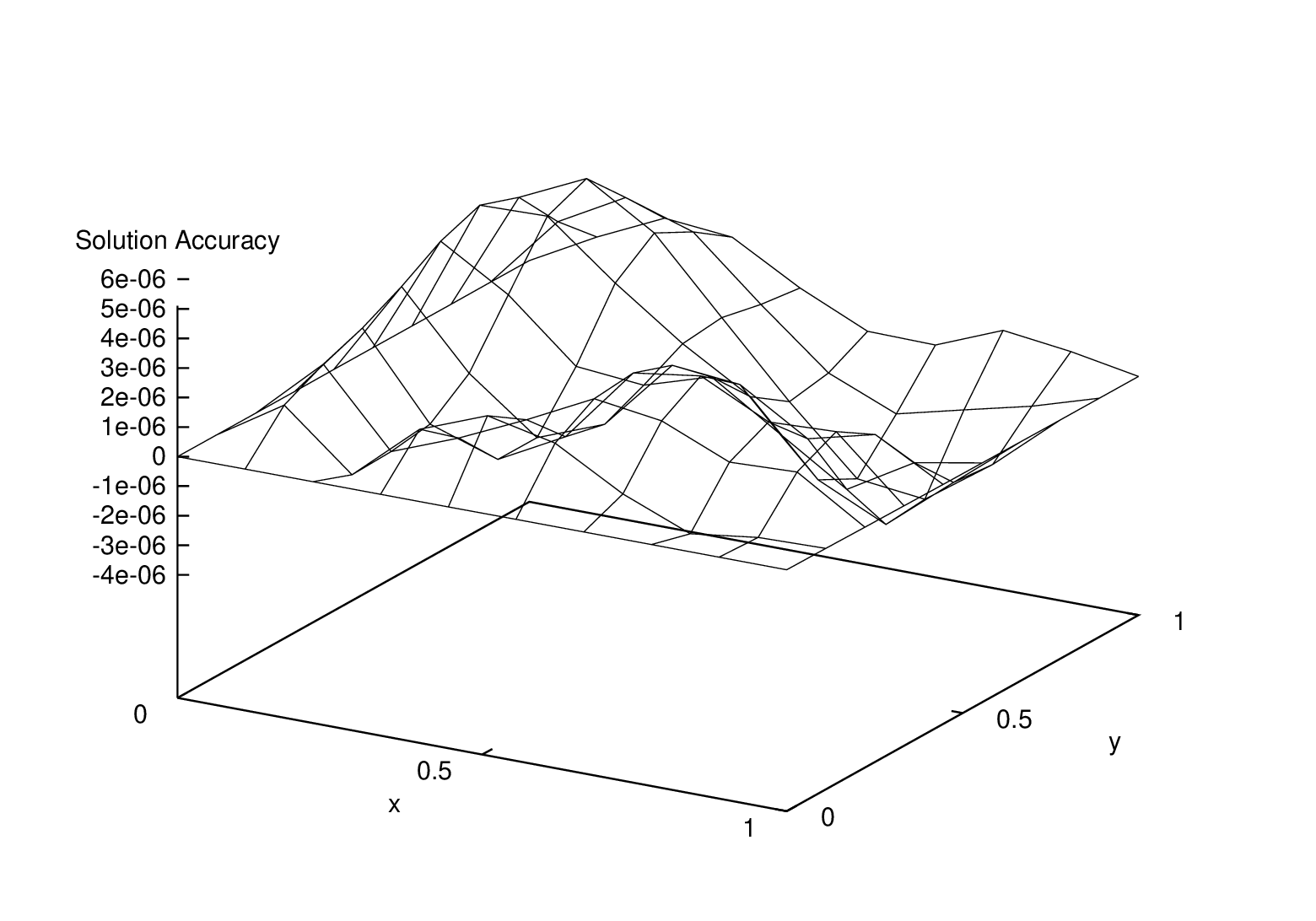}}
\caption{Problem 7: Accuracy of the computed solution at the training
points.}
\end{figure}

\begin{figure}
\centerline{\epsfysize=8cm\epsfxsize=13cm\epsffile{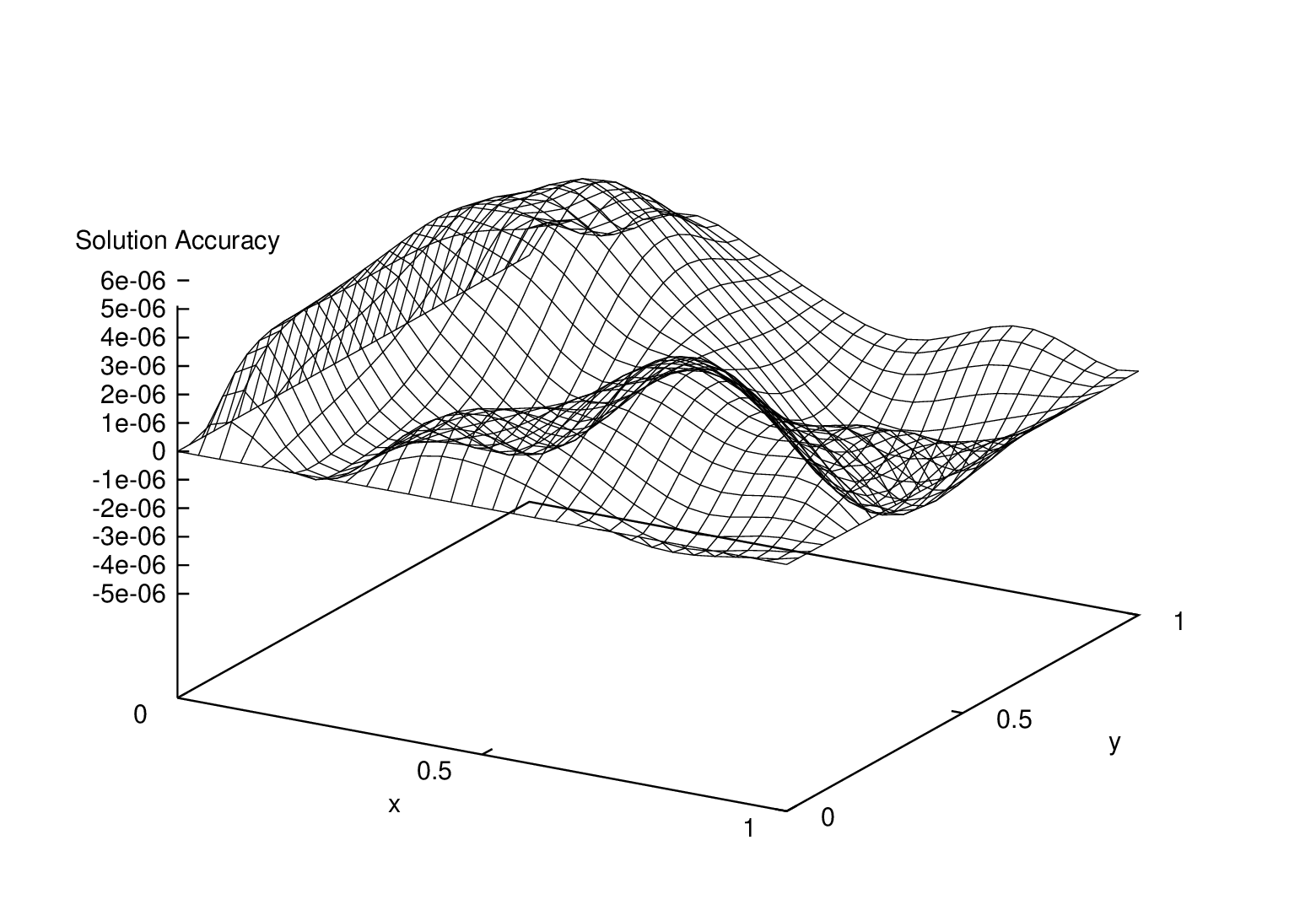}}
\caption{Problem 7: Accuracy of the computed solution at the test
points.}
\end{figure}

\subsection{Comparison with Finite Elements}

The above PDE problems were also solved with the finite element method which
has been widely acknowledged as one of the most effective approaches to the
solution of differential equations.
The characteristics of the finite element method
employed in this work are briefly summarized below. 
In the finite element approach the 
unknowns are expanded in piecewise continuous  
biquadratic elements \cite{FEM}:
\begin{equation}
\Psi=\sum_{i=1}^9 \Psi_i \Phi_i(\xi,n)
\end{equation}
where $\Phi_i$ is the biquadratic basis function and $\Psi_i$ is the 
unknown at the $i^{th}$ node of the element. The physical domain $(x,y)$
is mapped on the computational domain $(\xi, n)$ through the isoparametric
mapping:
\begin{equation}
x=\sum_{i=1}^9 x_i \Phi_i(\xi,n)
\end{equation}
\begin{equation}
y=\sum_{i=1}^9 y_i \Phi_i(\xi,n)
\end{equation}
where $\xi$ and $n$ are the local coordinates in the computational domain
$(0 \leq \xi, n \leq 1)$ and $x_i$, $y_i$ the $i^{th}$ node coordinates
in the physical domain for the mapped element.

The used Galerkin Finite Element Method (GFEM) calls for the weighted residuals
$R_i$ to vanish at each nodal position $i$:
\begin{equation}
R_i= \int_{D} G(x,y) det(J) d\xi dn = 0
\end{equation}
where $G$ is given by equation (1) 
and $J$ is the Jacobian of the isoparametric mapping:
\begin{equation}
det(J)=\frac{\partial x}{\partial \xi} \frac{\partial y}{\partial n}-
       \frac{\partial x}{\partial n} \frac{\partial y}{\partial \xi}
\end{equation}
This requirement  along with the imposed
boundary conditions constitute a set of nonlinear algebraic equations
($R_i=0$). The
inner products involved in the finite element formulation are computed using
the nine-node Gaussian quadrature. The system of equations is solved for the
nodal coefficients of the basis function expansion using the Newton's method
forming the Jacobian of the system explicitly  (for both linear and nonlinear
differential operators):
\begin{equation}
B \Delta \vec{\Psi}^{(n+1)}=-\vec{R}
\end{equation} 
\begin{equation}
\vec{\Psi}^{(n+1)}=\vec{\Psi}^{(n)} + \Delta \vec{\Psi}^{(n+1)}
\end{equation} 
where the superscript $n$ denotes the iteration number and 
$B$ is the global Jacobian of the system of equations $\vec{R}$:
\begin{equation}
B_{ij}=\frac{\partial R_i}{\partial \Psi_j}
\end{equation}
The initial guess
$\vec{\Psi}^{(0)}$ is chosen at random. For linear problems convergence 
is achieved in one iteration and for non-linear problems in 1-5 iterations.

All PDE problems 5-8 are solved on a rectangular domain of 18$\times$ 18 
elements resulting in a linear system with 1369 unknowns. 
This is in contrast 
with the neural approach which assumes a small number of parameters (30 for
ODEs and 40 for PDEs), but requires more sophisticated minimization
algorithms. As the number of employed elements increases the finite
element approach requires an excessive number of parameters. This fact  
may lead to memory requirements that exceed 
the available memory resources. 

\begin{table}
\begin{center}
\begin{tabular}{||c|c|c|c|c||} \cline{2-5}
     \multicolumn{1}{c|}{}
   & \multicolumn{2}{c|}{\em Neural Method}
   & \multicolumn{2}{c||}{\em Finite Element} \\
     \cline{1-5}
     \multicolumn{1}{||c|}{Problem No.}
   & \multicolumn{1}{c|}{Training set}
   & \multicolumn{1}{c|}{Interpolation set}
   & \multicolumn{1}{c|}{Training set}
   & \multicolumn{1}{c||}{Interpolation set} \\ \hline
  $5$  & $5\times10^{-7}$ &  $5\times 10^{-7}$  &$ 2\times 10^{-8}$ &$ 1.5\times 10^{-5}$    \\ \hline
  $6$  & $0.0015$ &  $0.0015$  & $0.0002$    & $0.0025$    \\ \hline
  $7$  & $6\times 10^{-6}$  &  $6\times 10^{-6}$  & $7\times 10^{-7}$ & $4\times 10^{-5}$    \\ \hline
  $8$  & $1.5\times 10^{-5}$  & $1.5 \times 10^{-5}$ & $6\times 10^{-7}$ & $4 \times 10^{-5}$    \\ \hline
\end{tabular}
\end{center}
\caption{Maximum deviation from the exact solution for the neural and the
finite element methods.}
\end{table}

In the finite element case, interpolation is
performed  using a rectangular grid 
of 23 $\times$ 23 equidistant points (test points). 
For each pair of nodal coordinates $(x, y)$ of this grid, 
we correspond a pair of
local coordinates $(\xi, n)$ of a certain element of the original 
grid where
we have performed the computations.
The interpolated
values are computed as:
\begin{equation}
\Psi (\xi, n)=\sum_{i=1}^9 \Psi_i \Phi_i(\xi,n)
\end{equation}
for the element that corresponds to the global coordinates $(x, y)$.
It is clear that the solution is not expressed in closed analyical form
as in the neural case, but additional computations are required in order to
find the value of the solution at an arbitrary point in the domain.
Figures 17-22 display the deviation $|\Psi(x,y)-\Psi_a(x,y)|$ 
for PDE problems 5-7 (figures concerning problem 8
are similar to those of problem 7). For each problem  
two figures are presented displaying  
the deviation at the training set and at the interpolation set of
points respectively.
Table 1 reports the maximum deviation corresponding to the neural 
and to the finite element method at the training and at the interpolation
set of points. 
It is obvious that at the training points the solution  
of the finite element method is very
satisfactory and in some cases it is better than that obtained
using the neural method. It is also clear that the 
accuracy at the interpolation points
is orders of magnitude lower as compared to that at  the training 
points. On the contrary, the neural method provides
solutions of excellent interpolation accuracy, since, as
Table 1 indicates, the deviations at the training and at the interpolation
points are comparable. It must also be stressed that the accuracy of the
finite element method decreases as the size of the grid becomes smaller, and
that the neural approach considers a mesh of 10$\times$10 points while
the in the finite  element case a 18$\times$18 mesh was employed.

\begin{figure}
\centerline{\epsfysize=8cm\epsfxsize=13cm\epsffile{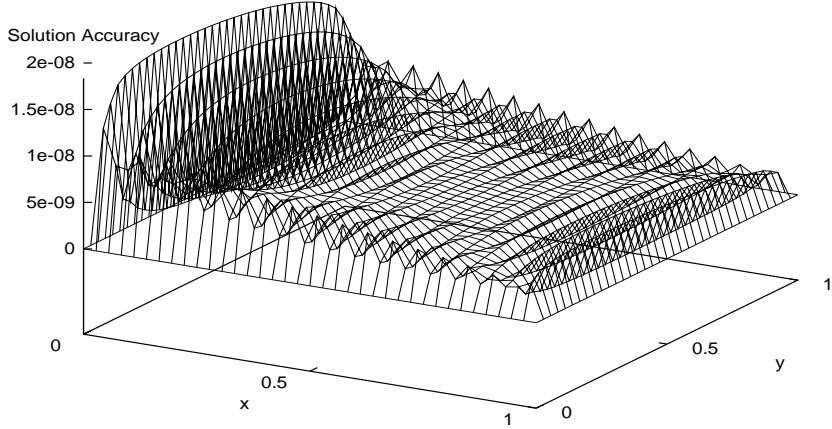}}
\caption{Problem 5: Accuracy of the FEM solution at the training points.}
\end{figure}

\begin{figure}
\centerline{\epsfysize=8cm\epsfxsize=13cm\epsffile{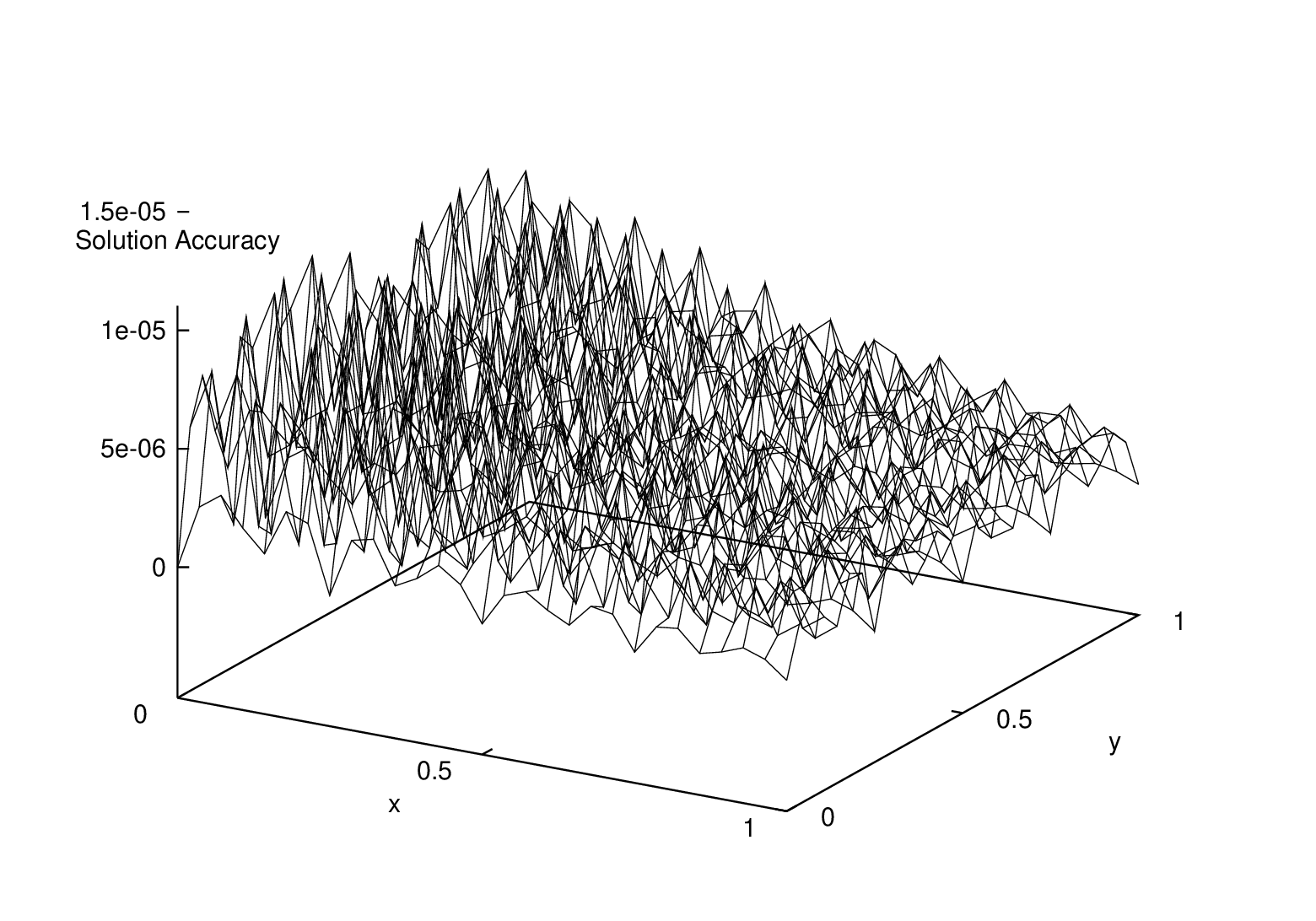}}
\caption{Problem 5: Accuracy of the FEM solution at the test points.}
\end{figure}

\begin{figure}
\centerline{\epsfysize=8cm\epsfxsize=13cm\epsffile{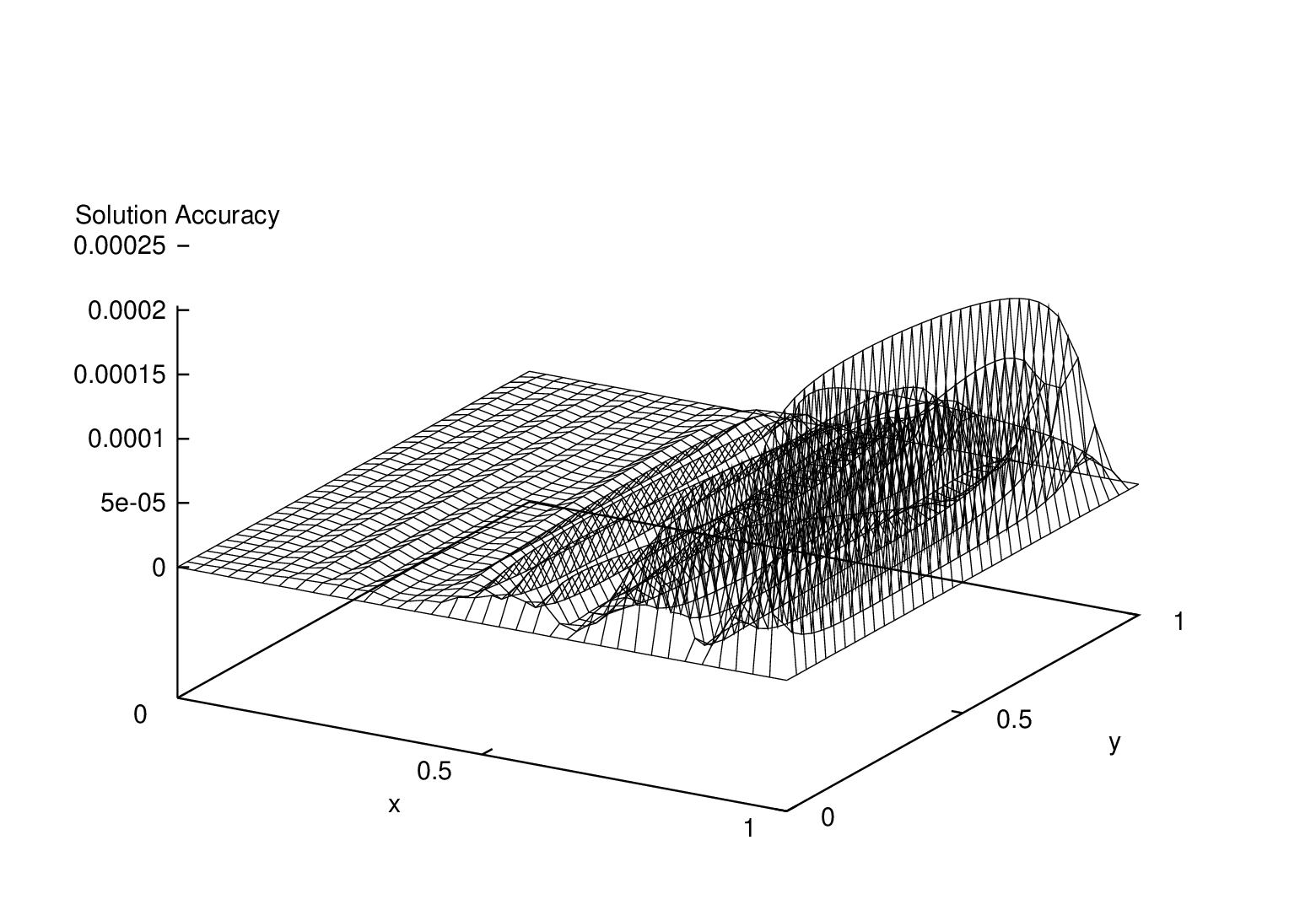}}
\caption{Problem 6: Accuracy of the FEM solution at the training points.}
\end{figure}

\begin{figure}
\centerline{\epsfysize=8cm\epsfxsize=13cm\epsffile{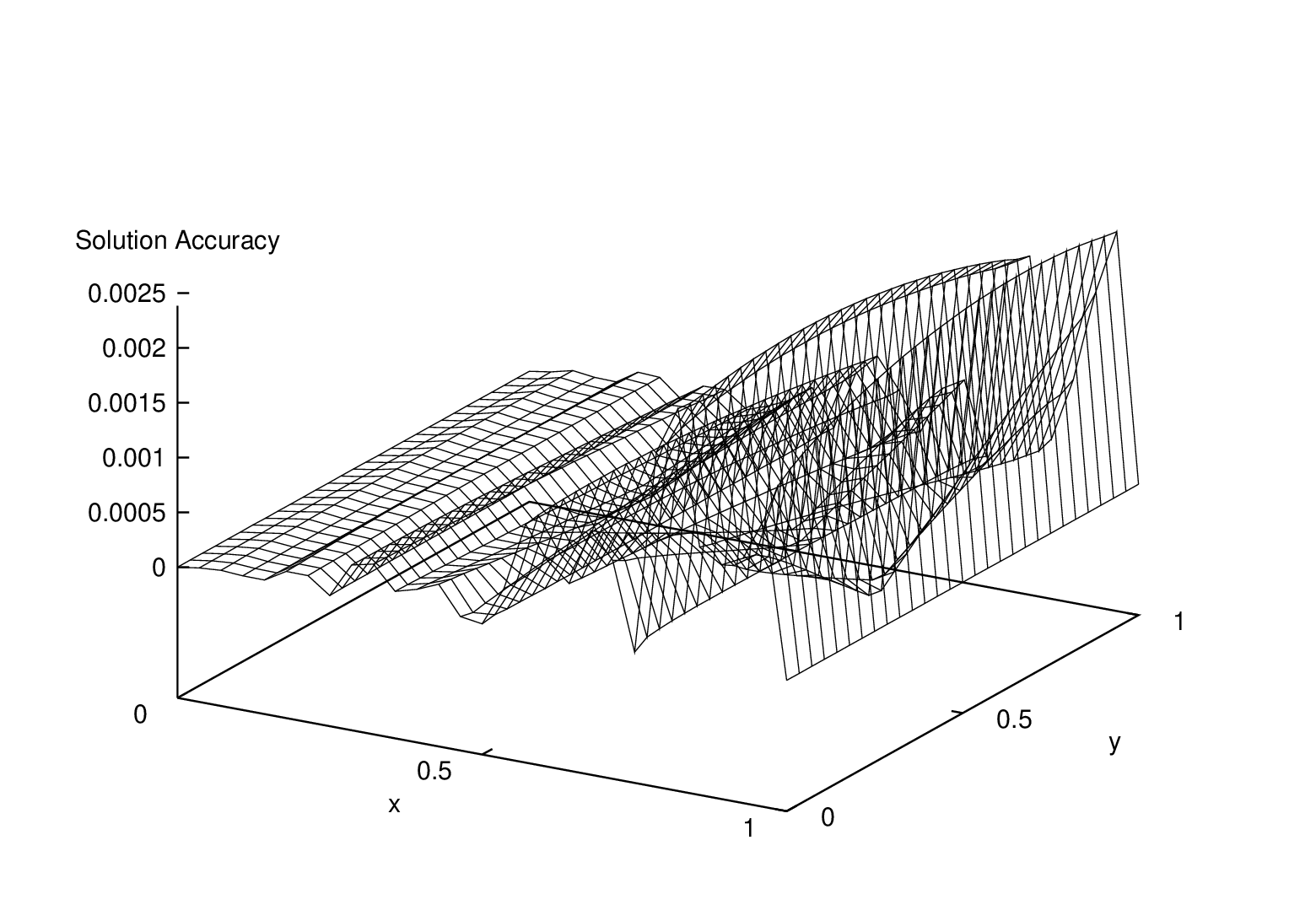}}
\caption{Problem 6: Accuracy of the FEM solution at the test points.}
\end{figure}

\begin{figure}
\centerline{\epsfysize=8cm\epsfxsize=13cm\epsffile{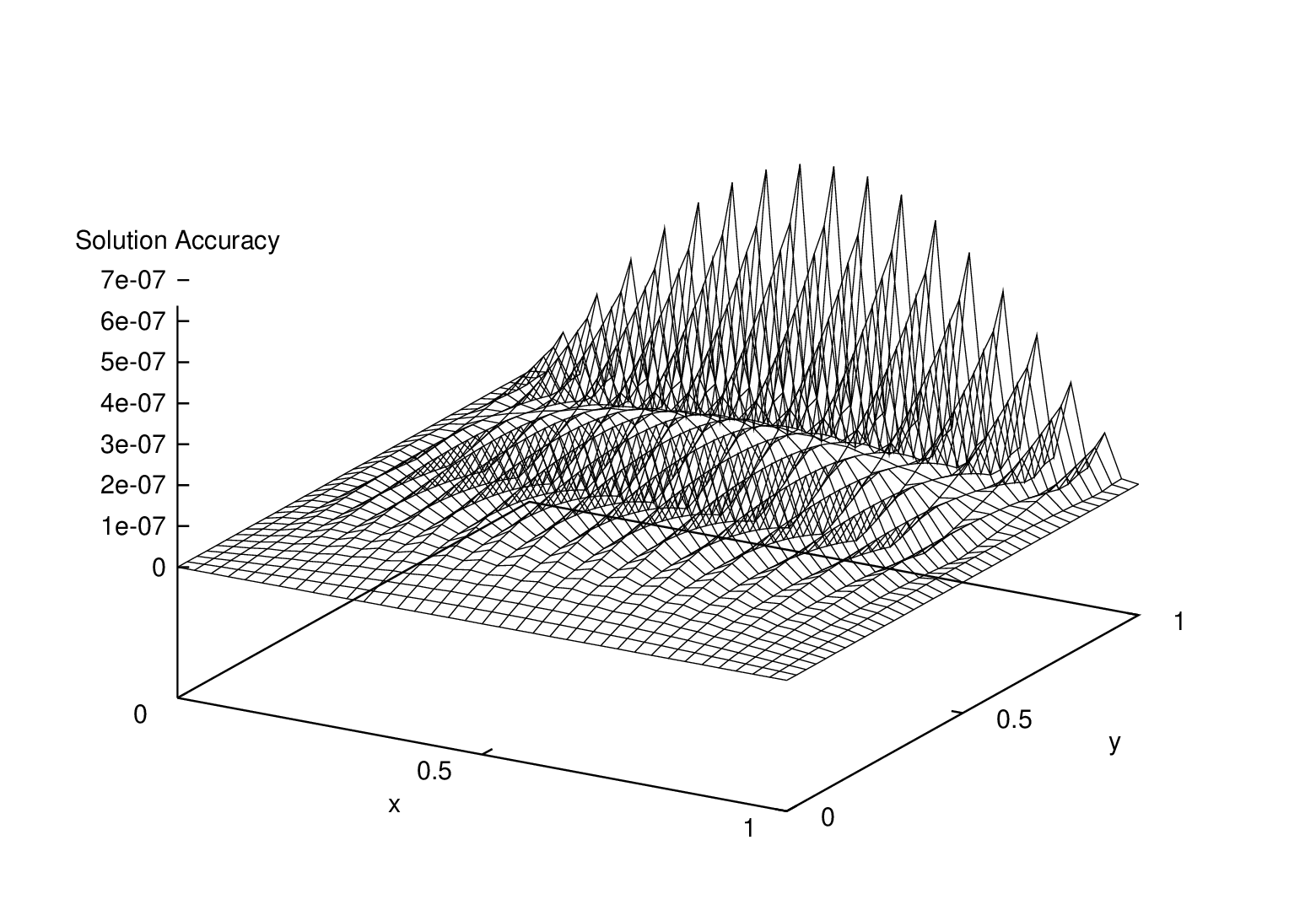}}
\caption{Problem 7: Accuracy of the FEM solution at the training points.}
\end{figure}

\begin{figure}
\centerline{\epsfysize=8cm\epsfxsize=13cm\epsffile{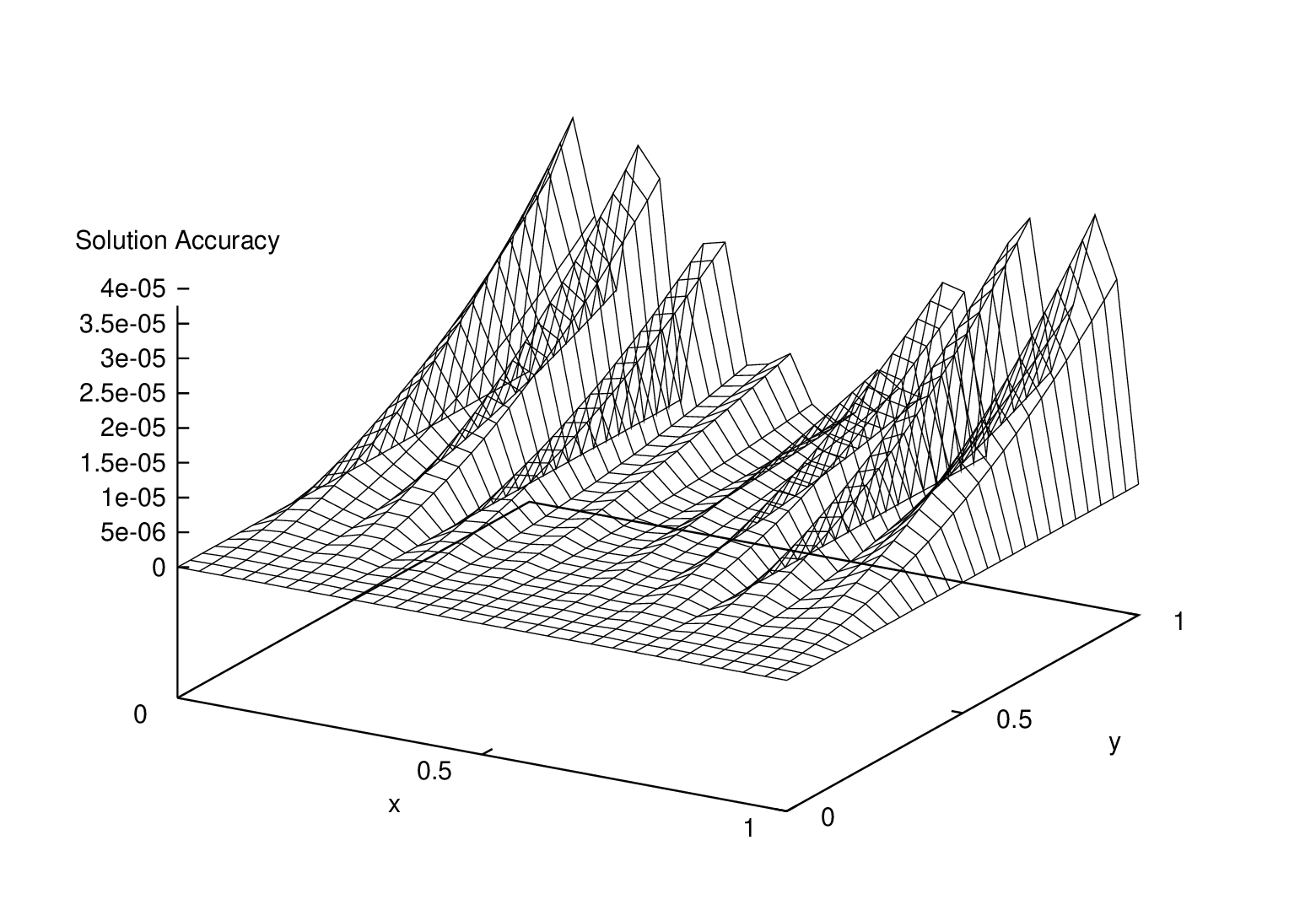}}
\caption{Problem 7: Accuracy of the FEM solution at the test points.}
\end{figure}

\section{Conclusions and Future Research}

A method has been presented for solving differential equations that relies
upon the function approximation capabilities of the
feedforward neural networks and provides
accurate and  differentiable solutions in a  closed  analytic form.
The success of the method can be attributed to two factors.
The first one is the employment of neural networks that are excellent
function approximators and the second is the form  
of the trial solution that satisfies  by construction
the BCs and therefore the constrained optimization problem becomes 
a substantially simpler unconstrained one.
 
Unlike most previous approaches, the method is general and can be applied
to both ODEs and PDEs by constructing the appropriate form of the trial 
solution. As indicated by our experiments the method exhibits 
excellent generalization performance since  the 
deviation at the test points was in no case greater than the maximum deviation 
at the training points. This is in contrast with the finite element
method where the deviation at the testing points
was significantly greater than the deviation at the training points.

We note  that the neural architecture employed 
was fixed in all the experiments and we did not 
attempt to find optimal configurations or to study the effect of the 
number of hidden units on the performance of the method. Moreover, we
did not consider architectures containing more than one hidden layers.
A study of the effect of the neural architecture on the quality    
of the solution constitutes one of our research objectives.

Another issue that needs to be examined is related with the sampling 
of the grid points that are used for training. In the above experiments
the grid was constructed in a simple way by considering equidistant points.
It is expected that better results will be obtained in the case where the
grid density will vary during training according to the corresponding
error values. This means that it is possible to consider more 
training points at regions where the error values are higher.   

It must also be stressed that the proposed method can easily be used for
dealing with domains of higher dimensions (three or more). As the 
dimensionality increases, the number of training points becomes large.
This fact becomes a serious problem for methods that consider local
functions around each grid point since the required number of parameters
becomes excessively large and, therefore, 
both memory and computation time requirements become intractable. In the
case of the neural method the number of training parameters remains 
almost fixed as the problem dimensionality increases. 
The  only effect on the computation time stems from the fact 
that each training pass requires the presentation of
more points, i.e. the training set becomes larger. This problem can be tackled  
by considering either parallel implementations, or implementations
on a neuroprocessor that can be embedded in a conventional machine and
provide considerably better execution times. Such an implementation on neural  
hardware is one of our near future objectives, since it will permit 
the treatment of  many difficult real world problems. 
Finally we aim at extending the approach to treat other
problems of similar nature, as for example eigenvalue problems
for differential operators.

One of us (I. E. L.) acknowledges
partial support from the General
Secretariat of Research  and  Technology under contract
PENED 91 ED 959.

\end{document}